\documentclass[aps,showpacs,floatfix,notitlepage,superscriptaddress,longbibliography]{revtex4-1}
\usepackage{epsfig,graphics,amssymb,amsmath,subeqnarray,color,bm,bbm}
\usepackage[toc,page]{appendix}
\usepackage{mathtools}
\usepackage{xcolor}
\usepackage{float}
\usepackage{graphicx}
\usepackage[colorlinks=true,allcolors=blue]{hyperref}

\begin{document}
\title{Enhanced diffusion of a tracer particle in a lattice model of a crowded active system}
\author{Leila Abbaspour}
\email{Leila.abbaspour@theorie.physik.uni-goettingen.de}
\author{Stefan Klumpp}
\email{Stefan.klumpp@phys.uni-goettingen.de}
\affiliation{Institute for the Dynamics of Complex Systems, Georg August University of G{\"o}ttingen, Friedrich-Hund-Platz 1, 37077 G{\"o}ttingen, Germany.}
\date{\today}


\begin{abstract}
Living systems at the sub-cellular, cellular, and multi-cellular level are often crowded systems that contain active particles. The active motion of these particles can also propel passive particles, which typically results in enhanced effective diffusion of the passive particles. Here we study the diffusion of a passive tracer particle in such a dense system of active crowders using a minimal lattice model incorporating particles pushing each other. We show that the model exhibits several regimes of motility and quantify the enhanced diffusion as a function of density and activity of the active crowders. Moreover, we demonstrate an interplay of tracer diffusion and clustering of active particles, which suppresses the enhanced diffusion. Simulations of mixtures of passive and active crowders  show that a rather small fraction of active particles is sufficient for the observation of enhanced diffusion. 
\end{abstract}

\maketitle
\section{Introduction}
The behavior of individual particles within a crowded environment is influenced by the complex collective behavior of the surrounding particles \cite{ellis2001,Dix2008,Bechinger2016,klumpp2019}. Indeed, many biological systems are densely packed. Molecular systems in the cytoplasm of cells \cite{ellis2001,Zimmerman1993}, colonies of bacteria \cite{Mendelson1999}, bio-films \cite{Hall2004}, and tissues \cite{Salbreux2017} are a few examples. Often, the crowded environment contains active particles that  are driven out of equilibrium due to the local injection and dissipation of energy. On the molecular scale these active particles may be molecular motors pulling cargoes through cytoplasm and 'stirring it' \cite{MacKintosh2007,Klumpp2008,Korn2009} or enzymes exhibiting enhanced diffusion \cite{Ramin2009,Illien2017}, on the cellular scale, these could be self-propelled cells swimming in a suspension or moving through a tissue or a colony of cells \cite{Sokolov2012,Dunkel2013}. This means that detailed balance is not fulfilled in these systems, so they cannot be described by equilibrium statistical mechanics.

The observation of individual 'probe' or 'tracer' particles within such a system provides a method to quantify crowding effects as well as to probe for the degree of crowding if this is difficult to determine directly. Such probes include the diffusion of the tracer particle \cite{konopka2009,parry2014} as well as specific chemical reactions or conformational changes such as the compaction of a polymer chain as measured by FRET-probes at the two ends of the polymer \cite{boersma2015,gnutt2015}.

Diffusion of molecules and particles in complex environments has been studied on a range of length scales and the complex environment can give rise to dynamic phenomena such as anomalous diffusion \cite{koch1988,Ghosh2016}. 
Specifically, diffusion of a passive tracer particle in an active  system has been studied in various experimental and theoretical model systems.
Notably, Wu and Libchaber \cite{Wu2000} studied the trajectories of polystyrene beads  in a suspension of swimming bacteria and observed enhanced diffusion of the beads. 

Such enhanced diffusion has been studied extensively, both theoretically \cite{Hugues2001,Zaid2011,valeriani2011,Pushkin2013,Ramin2014,Kasyap2014,Morozov2014,Yang2020} and experimentally \cite{Caspi2000,Kim2004,Leptos2009,valeriani2011,Mino2011,Mino2013,Jepson2013,Koumakis2013}.
Extensions of the simplest system include  $3$ dimensional systems \cite{Jepson2013}, the presence of obstacles  \cite{Mino2011} and confinement, and effective interaction between passive particles \cite{Angelani2011}.

Interactions between particles in suspension can be complex and generally involve hydrodynamic interactions in addition to simple excluded volume and particle-particle collisions. Therefore, simplified minimal models systems can help to advance our understanding of the role of active processes. Both continuum models  \cite{Ghosh2015} and lattice models \cite{benichou2018,mejia2020} have been used for that purpose. Enhanced diffusion of a tracer in a crowded active system has been observed in models with and without hydrodynamic interactions \cite{valeriani2011,brady2017}. 
Following this approach, we study a minimal lattice model incorporating particle-particle collisions as well as different diffusion coefficients of different particle types to investigate the effect of activity and density of crowders on tracer diffusion. We quantify the enhanced diffusion as a function of these parameters and relate the observed tracer diffusion to clustering of active particles. Finally, we generalize our analysis to the case of a binary mixture of active and passive particles. Our results indicate how the interplay between activity and density of crowders can lead to significant enhancement of the diffusion of a tracer, but also limits that enhancement through clustering of the active particles.

\begin{figure}
    \centering
   \includegraphics[width=150mm,scale=1.]{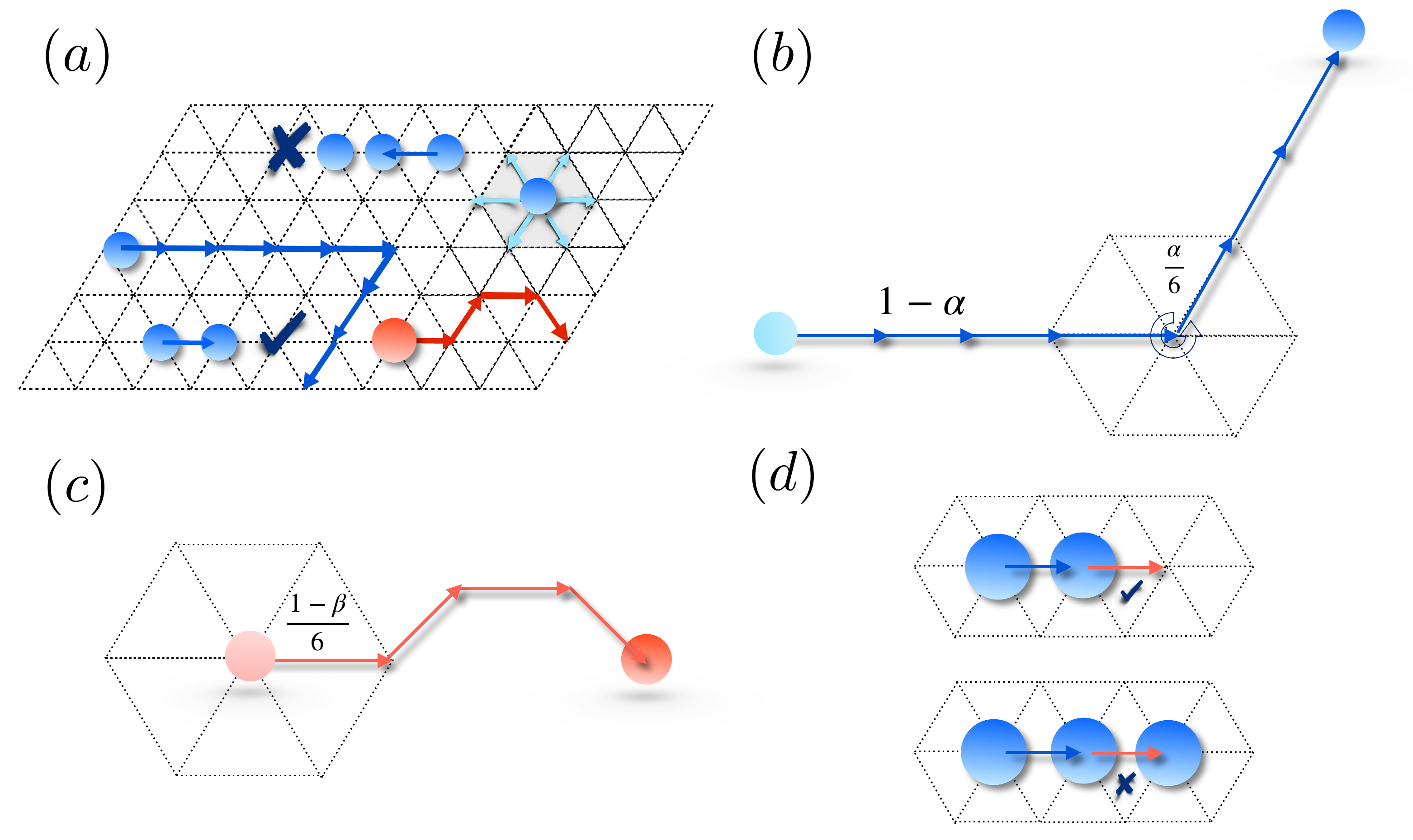}
    \caption{ Lattice model of a dense system of active particles (blue) containing a  passive tracer particle (red).  a) Overview of the model on a hexagonal lattice. b) Active particles perform run and tumble motion and move persistently in the same direction with probability $1-\alpha$ and randomize that direction (tumble) with  probability $\alpha$ (i.e., $\alpha/6$ for each direction). c) The passive tracer particle (red) performs a simple random walk, its  diffusion coefficient is modulated by the parameter $\beta$. d) Collision rules among particles.
    }
    \label{Model}
\end{figure}
\section{\label{sec:level1}Model}

We consider a hexagonal lattice (see Fig.~\ref{Model}(a)) in two dimensions, consisting of $ M = 256 \times 256$ sites, with periodic boundary conditions. There are in total $N$ particles on the lattice, of which $N_a=\chi_A N$ are active and can self-propel following a simplified run and tumble motion, and the rest are passive and exhibit diffusive behaviour. We define $\rho$ as the number density of the system, which is defined as the total number of particles divided by the number of lattice sites.

For updating the position of all particles  at each time step, $N$ random actions take place in the system. A particle may be picked more than once in one single time step, giving rise to fluctuation of its speed. On average, however, all particles are picked equally often.

The active particles have run-and-tumble dynamics, where they move persistently in one direction (initially chosen at random), and then reorient with a tumbling probability $\alpha$, resulting in long directed runs with abrupt turns. At each time step, the active particle moves in the same direction as in its last step with probability $1- \alpha$, and tumbles to choose a new direction with probability  $\alpha / 6$. For instance, for $\alpha =0.001$, the active particle takes, on average, $1000$ persistent steps in the same direction, and then tumbles and chooses one out of the six possible directions with equal probability (see Fig.~\ref{Model}(b)). In the case of passive particles we set $\alpha=1$: the particle reorients at each time step and therefore steps to each site with a probability of $1/6$. Since our aim is to study the effect of active particles on the overall motion of a passive particle, we put (at least) one passive `tracer' in the system, which in general has a different Brownian diffusion coefficient set by a parameter $\beta$. The tracer steps to each neighbour site with probability $(1-\beta)/6$, with $0 \leq \beta \leq 1$ as shown in Fig.~\ref{Model}(c). When $\beta=0$ the tracer has the same dynamics as the passive crowders. As $\beta$ increases, the tracer diffuses more slowly and in the limiting case of $\beta = 1$, it can only move by being pushed by its neighbors.

We now define collision rules among the particles. We use a modified exclusion rule, which in addition to ensuring that each site is occupied by at most one particle also allows for particles pushing each other. In each stepping attempt of a particle, we check whether the target site on the lattice is empty or occupied.  If it is empty, then the particle is allowed to update its position, as described above. If the site is occupied, then there are two possibilities depending on the second neighbor site of the particle in the same direction. If the latter site is empty, the stepping particle pushes the particle in the target site to the second neighbor site and moves to the target site itself. Otherwise, the update is rejected (see Fig.~\ref{Model}(d)). Note that there is no alignment interaction between particles in the pushing process. All simulations described in the following were run for $10^6$ time steps with $1000$ realizations of each condition.


\section{Results}


\subsection{Tracer in a purely active system}

\subsubsection{Mean square displacement}



\begin{figure*}
    \centering 
    
   \includegraphics[width=1.\textwidth]{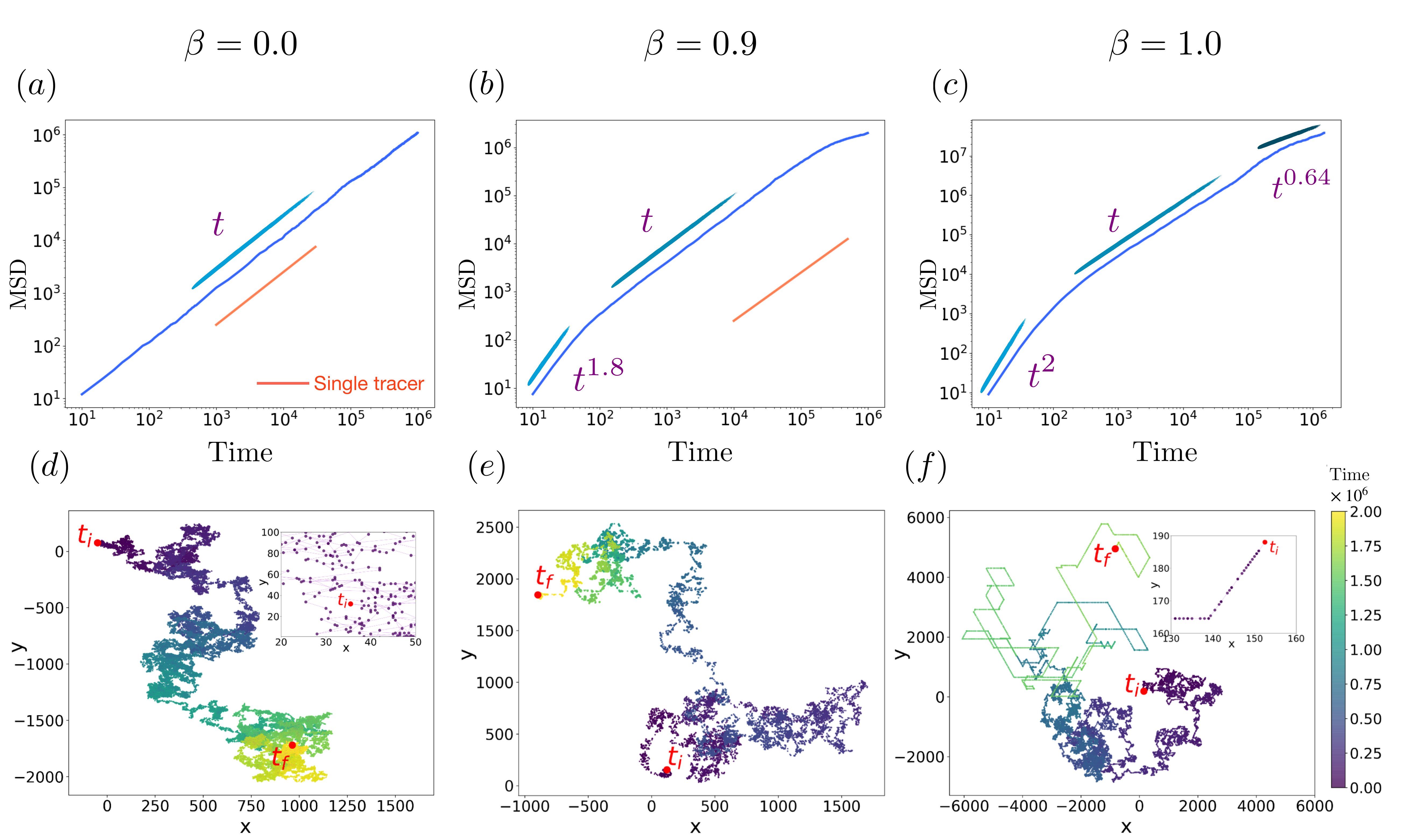}
    \caption{Diffusion of tracer particles with different diffusion coefficients in a system of active crowders.  (a-c) The mean square displacement shows diffusive motion for $\beta=0$ and anomalous diffusion for $\beta=0.9,1$. The red lines indicate the diffusive motion of the tracer in an otherwise empty lattice, in (c) the tracer would not move in that case. (d-f) show trajectories, the insets show the trajectories at the early stages. Each column corresponds to the value of $\beta$ indicated above it. The other parameters are $\rho=0.025$, $\alpha=0.0001$.}
    \label{MSD}
\end{figure*}

We first discuss the results for a passive tracer particle in the presence of only active crowders (i.e., $N_a =N$ or $\chi_A \approx 1$). We considered densities in the range of $0.01 \leq \rho  \leq 0.4$ and the tumbling probability of active particles between $0.0001 \leq \alpha < 1$. 

We begin our analysis by evaluating the mean square displacement (MSD) of the tracer particles. We note that for Brownian motion, the MSD is given by,
$\left< \left( \Delta r \right)^2 \right> = 2 d D t$, where $\Delta r$ is the displacement of a particle in a given time interval of $t$, $d$ is the spatial dimension (which is $2$ in our case), $D$ is the diffusion coefficient, and $\left<\cdot\right>$  denotes an ensemble average. So-called anomalous diffusion is characterised by an MSD $ \left< \left( \Delta r \right)^2 \right> \sim t^\gamma$, where $\gamma \neq 1$ is a real positive number that classifies the different types of diffusion: sub-diffusion for $ 0 < \gamma <1$ and, super-diffusion for $ \gamma > 1 $ (where $\gamma = 2$ gives rise to ballistic motion).

In Figs.~\ref{MSD}(a)-(f), we have plotted the MSD and corresponding trajectories as a function of time for the three cases of $\beta=0$, $0.9$, and $1$. In all three cases, the density and the tumbling probability of crowders are the same: $\rho=0.025$ and $\alpha=0.0001$. For a purely passive Brownian tracer with $\beta = 0$ (Figs.~\ref{MSD}(a) and (d)), the tracer shows normal diffusion, with an MSD that is linear in time throughout the simulation. The trajectory is erratic at both long and short time (Figs.~\ref{MSD}(d) and its inset, respectively).

It is useful to compare the diffusion coefficient of the tracer in this system to that  of a passive tracer in the absence of active crowders. For a single passive particle on an isotropic lattice, the diffusion coefficient is given by $ D_0 = ( \kappa / 2d ) a^2 \lambda$, where $a$ is the lattice constant, $\kappa$ is the coordination number (the number of neighbour sites allowed for a jump), and $\lambda$ is the stepping rate to the nearest neighbour site \cite{braun1998}. In our model $a=1$, $\kappa=6$, $d=2$, and $\lambda = \frac{1-\beta}{6}$, leading to
\begin{equation}
 D_0 = \frac{1-\beta}{4}.
\end{equation}
Thus, for $\beta=0$ we find $D_0=1/4$, which as shown in Fig.~\ref{MSD}(a) (red line) indicates that the diffusion coefficient of our tracer particle is enhanced due to presence of active crowders.

\begin{figure}
    \centering
   \includegraphics[width=\linewidth]{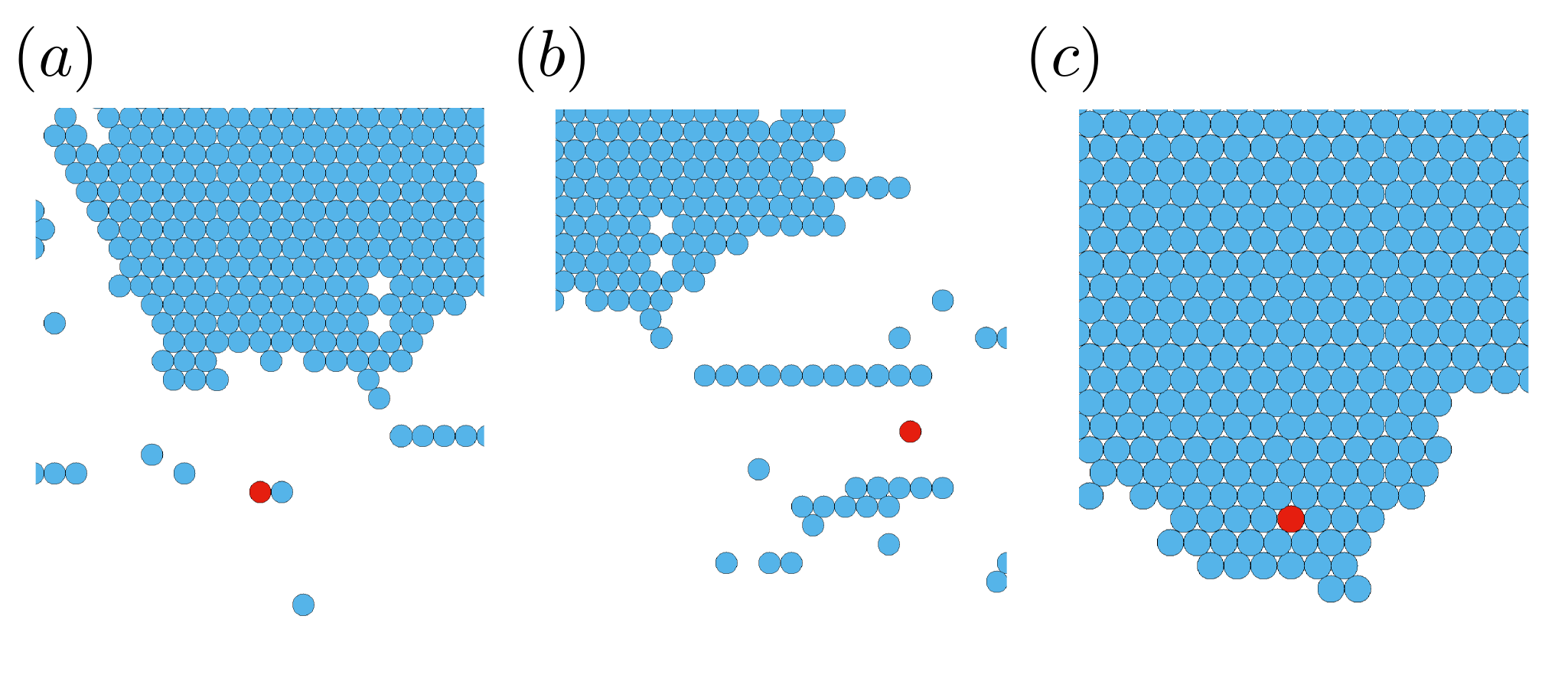}
    \caption{Snapshots of the simulation for  $\rho=0.1$ and $\alpha=0.0001$: the active particles are shown in blue and the tracer particle in red. a) The tracer (red) is next to an active particle (blue), which pushes it. b) The tracer is trapped in empty space. c) The tracer is trapped by active particles.}
    \label{Snapshot}
\end{figure}

For $\beta=0.9$ and $\beta=1$, by contrast, we can identify several diffusive regimes. A slow tracer with $\beta=0.9$ initially undergoes super-diffusion (Fig.~\ref{MSD}(b)) as it remains in front of an active particle over an extended period of time (see Fig.~\ref{Snapshot}(a)). The tracer inherits the direction of an individual active particle which drives it forward. This effect can also be seen in the inset of Fig.~\ref{MSD}(e), in which this directed motion yields a straight line. The observed diffusion exponent $\gamma$ at short time is found to be $1.8$, which is close to the ballistic regime with $\gamma=2$. After a longer time, diffusive motion is recovered, with erratic trajectories similar to those for $\beta=0$.

Finally, Figs.~\ref{MSD}(c) and (f) show the case where the passive tracer does not move by itself and only moves when pushed by an adjacent active particle. Initially, the tracer fully inherits the direction and the speed of the active particle that pushes it (Fig.~\ref{MSD}(f) inset). Since there is only self-propulsion, $\gamma$ is exactly $2$. For an intermediate time, diffusive behaviour is recovered. Then after some time, the motion becomes remarkably less random, exhibiting long directed paths and pauses, as depicted in the trajectory in Fig.~\ref{MSD}(f). Correspondingly, the exponent of the MSD is reduced to $\gamma \sim 0.64$, indicating a sub-diffusive behaviour. Inspection of snapshots of the lattice (Figs.~\ref{Snapshot}(b) and (c)) show that during the pauses, the tracer is either trapped in an empty region of the lattice and thus cannot be moved, or in a cluster of active particles, within which it is also immobilized. This observation indicates that, in this case, the environment in which the tracer diffuses through changes from a homogeneous environment, to coexistence of a dilute phase and clusters of active particles. Cluster formation will be discussed in more detail below.


\begin{figure*}
 \centering
   \includegraphics[width=\textwidth]{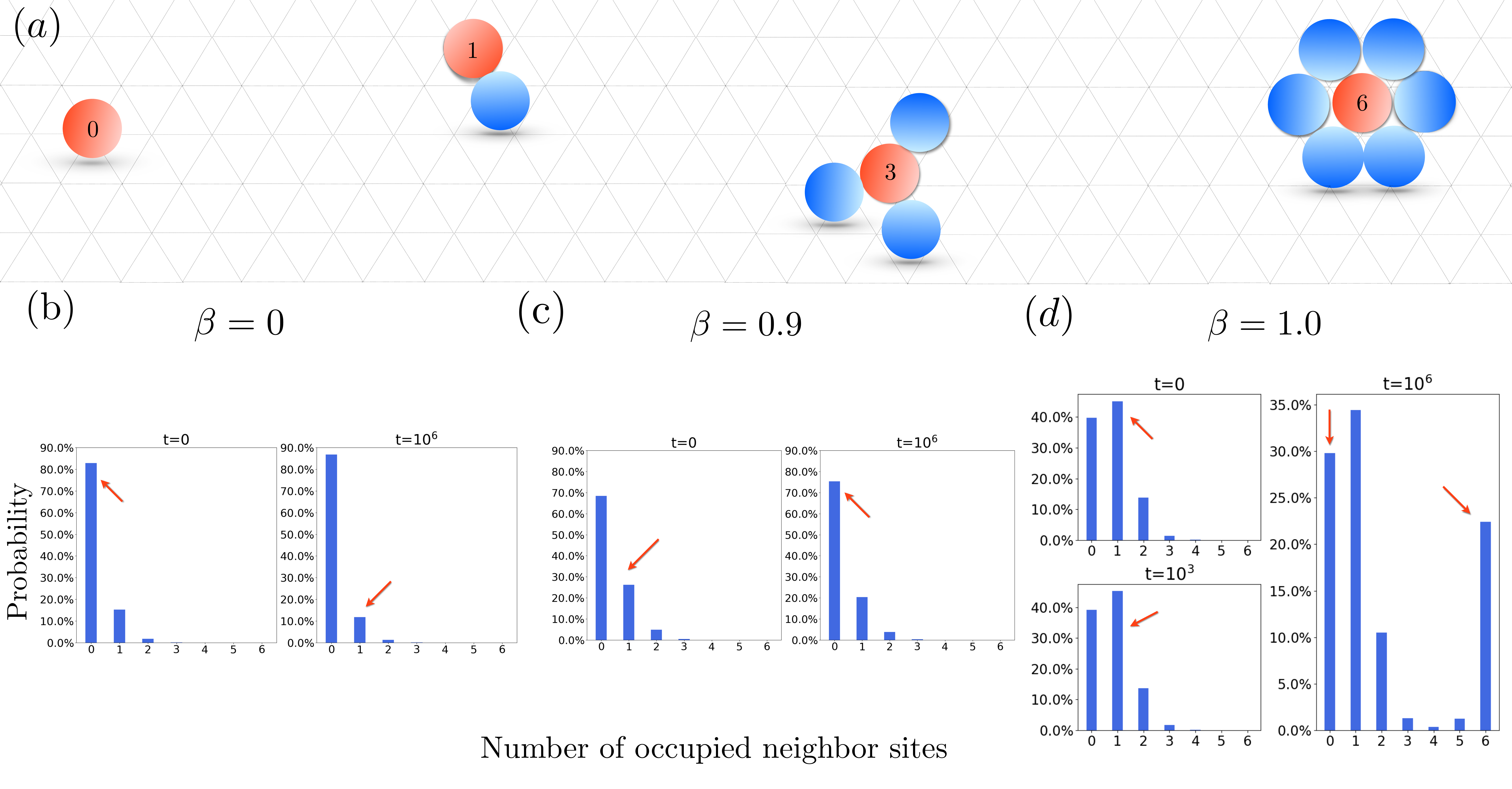}
    \caption{Trapping analysis. a) The sketch shows the tracer (red)  surrounded by different number of  active particles (blue). (b), (c) and (d) show the probability that $N_n$, neighbor sites are occupied (for $n=0, \dots 6$). The red arrows indicate the numbers of occupied neighbors most relevant for the different regimes in the MSD plots in Fig.~\ref{MSD}.}
    \label{Trap}
\end{figure*}


\subsubsection{Neighbour site occupation and trapping}

To quantify the effect of trapping on the motion of a tracer, we define a trapping parameter $N_n$, which characterizes the occupation of the six nearest neighbour sites of the tracer (see Fig.~\ref{Trap} (a)). Thus, $0\leq N_n\leq6$, with $N_n=0$ indicating that the tracer has no neighbouring particle and $N_n=6$ is the fully trapped state. 
When $N_n=0$, the tracer with $\beta =1$ is essentially trapped by the empty space and becomes immobile, while tracers with $\beta =0$ and $ \beta = 0.9 $ diffuse on their own. In the fully occupied state ($N_n = 6$), the tracer is immobile for all the cases. $N_n = 1$ results in persistent directed motion of the tracer if the neighboring active particle moves to the site of the tracer and $1-\beta<1-\alpha$, as the active particle then pushes the tracer for multiple steps. 

For the two cases $\beta =0$ and $ \beta = 0.9 $, the histograms of $N_n$ do not vary significantly over time, with the probability of having no neighbors being the largest (Figs.~\ref{Trap}(b) and (c)). As expected, the probability of having a single neighbor ($N_n=1$) is higher for $\beta=0.9$ than $\beta=0$, in agreement with the observed directed motion. For $\beta = 1$, $N_n = 1$ is the most probable configuration at all times. However, the histogram remarkably broadens over time  and becomes bi-modal at long times, with $N_n=0$ and $N_n=6$ occurring (almost) equally  frequently. This means that the three most likely configurations are those that lead to directed motion ($N_n=1$), trapping in empty space ($N_n=0$) or trapping in a cluster ($N_n=6$), which together explain the range of motion observed in Fig.~\ref{MSD}(f). 

\subsubsection{Effective diffusion coefficient at large times}

\begin{figure}
    \centering
   \includegraphics[width=0.8\linewidth]{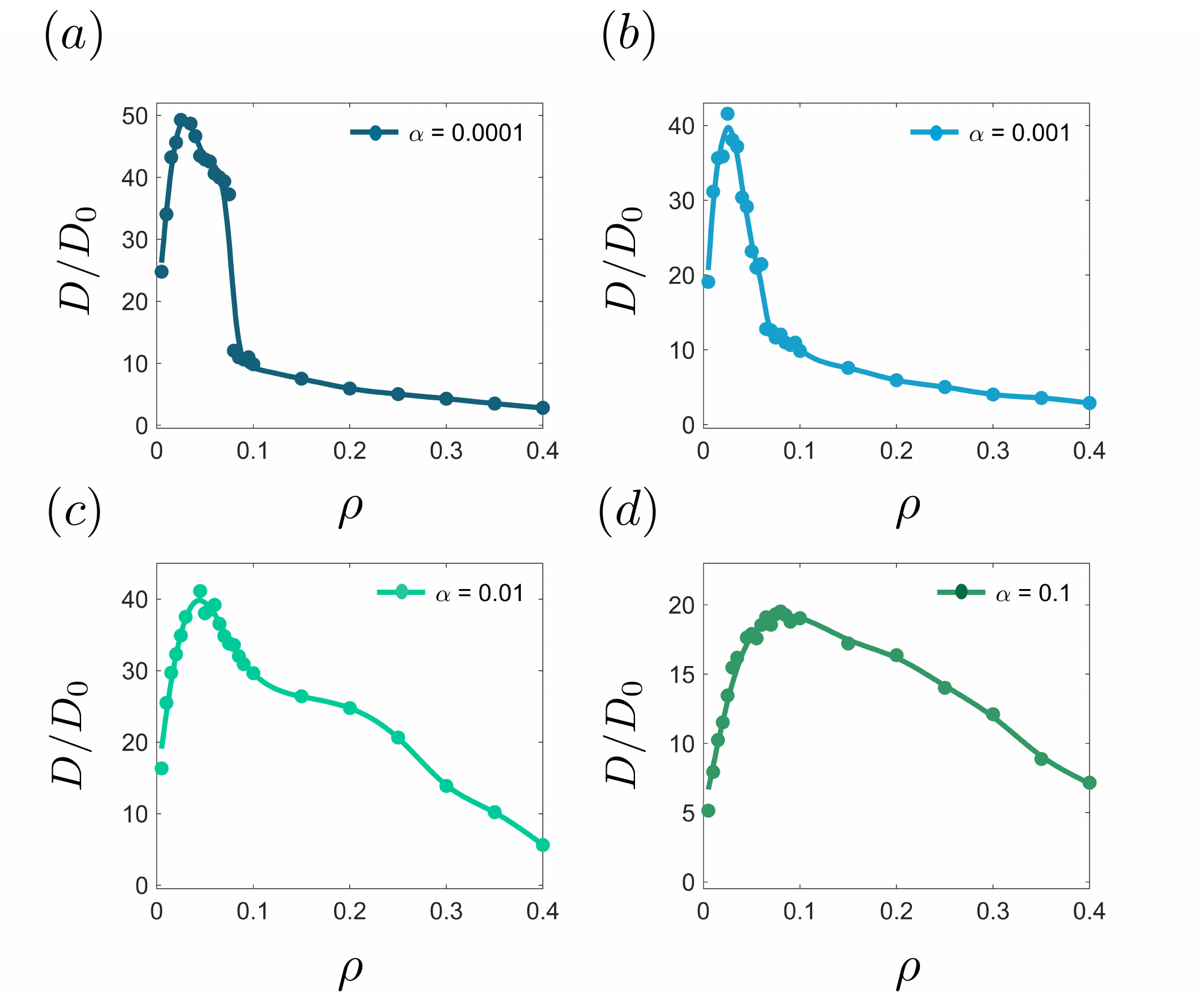}
    \caption{The relative diffusion coefficient of a tracer with $\beta=0.9$, normalized to  the diffusion constant $D_0$ of an isolated tracer.}
    \label{Diffusion}
\end{figure}

We have observed above that diffusion of the tracer is enhanced in the bath of active particles compared to the tracer on its own. To study this further, in Fig.~\ref{Diffusion}, we plot 
the diffusion coefficient at large times as a function of density, focusing on the case $\beta=0.9$. The simulation was performed for a system with $\chi_A=1$, and different values of $\alpha$. Remarkably, the effective diffusion coefficient has a $50$-fold increase when in the active bath, compared to the diffusion coefficient of an isolated tracer. Even at a low density ($\rho \approx 0.01$) of active crowders, the diffusion coefficient of the tracer is substantially greater than in the absence of active crowders, implying a strong effect of being pushed by active particles. 

Since the diffusion coefficient of a single passive tracer is a small constant in the absence of crowders and goes to  zero in the presence of a high density of crowders, the relative diffusion coefficient is expected to be a non-monotonic function of the  density. Furthermore, the form of this dependence should also vary with the tumbling probability. The plots in Fig.~\ref{Diffusion} show a clear density dependence of the diffusion coefficient. The diffusion coefficient is sharply enhanced when the density of active particles is increased from zero, then decreases after a pronounced peak. Clearly, the density of the system greatly affects the diffusion coefficient of the tracer. To further investigate this effect, we come back to the observation of  cluster formation. 

\subsubsection{Cluster formation}
\begin{figure}
   \centering
   \includegraphics[width=0.9\textwidth]{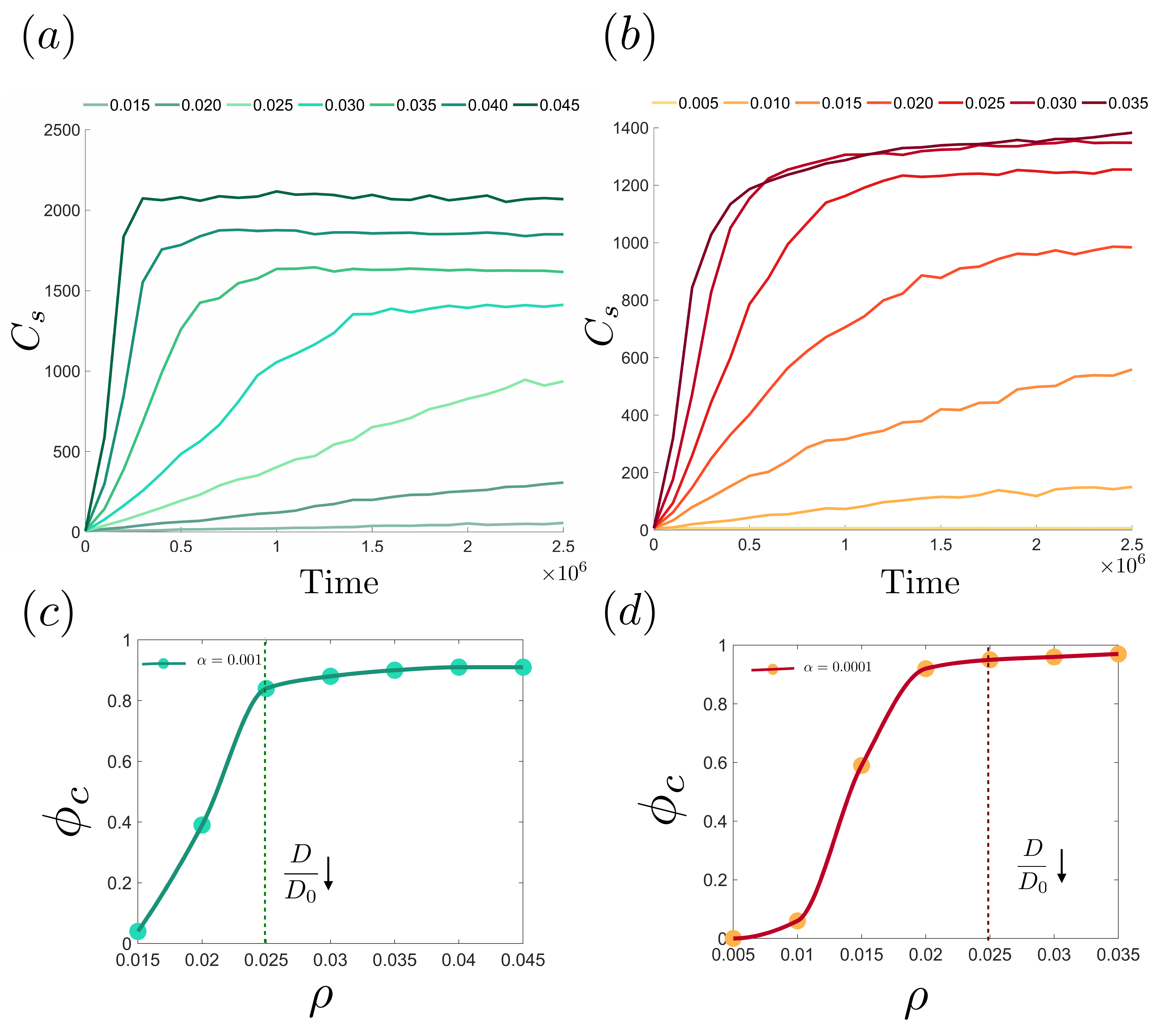}
    \caption{Cluster analysis: a) and b) Average cluster size as a function of time (averaged over 1000 realizations) for different densities. c) and d)  Fraction of particles assembled in clusters in the last snap shot of simulation for the same densities. The dashed lines mark the density in which the tracer exhibits the maximum diffusion coefficient. a,c) $\alpha=0.001$, b,d) $\alpha=0.0001$.
     }
    \label{cluster}
\end{figure} 

To analyze clustering, we define a cluster as a collection of at least six particles 'sticking together', i.e. being connected via mutual nearest neighbors. Particles without nearest neighbor or connected to fewer particles are considered as belonging to the gas phase. Clusters are identified with the data analysis framework
Freud \cite{freud}. We measured the average cluster size, $C_s$, as the average number of particles in a cluster over $1000$ different realizations of the simulation. Figs.~\ref{cluster}(a) and (b) show the time evolution of $C_s$ for different densities and for $\alpha=0.001$ and $0.0001$, respectively. As can be seen in Figs.~\ref{cluster}(a) and (b), cluster formation is accelerated by increasing the density of crowders.

In Fig.~\ref{cluster}(c) and (d), we plotted the fraction of particles detected within clusters, $\phi_c$, as seen in the last snapshot of the simulation which provides an estimate of the fraction of particles assembled in the clusters. $\phi_c$ can be described by $(N_c \times C_s) / N$, where $N_c$ is the average number of clusters. As a function of the density, $\phi_c$ shows a clear transition from a state in which a very small fraction of particles are in the clusters to a state where almost all particles are in clusters. 

To relate clustering to the tracer diffusion discussed before, we indicate with the vertical dashed lines  the density at which the tracer's  diffusion coefficient for the given tumbling probability is maximal. These lines fall near the transition to global clustering, suggesting an intimate relation of the decline in diffusion to the onset of phase separation. When the active crowders start to form clusters, there is coexistence of a dilute gas phase and clusters of active particles and the probability of active crowders being in the dilute phase  decreases as long as the cluster size increases. Consequently, the diffusion coefficient of the tracer decreases as there are fewer and fewer active crowders in the dilute phase to push the tracer and enhance its diffusion.


\subsection{Binary Mixture}

\begin{figure}
   \centering
   \includegraphics[width=0.75\linewidth]{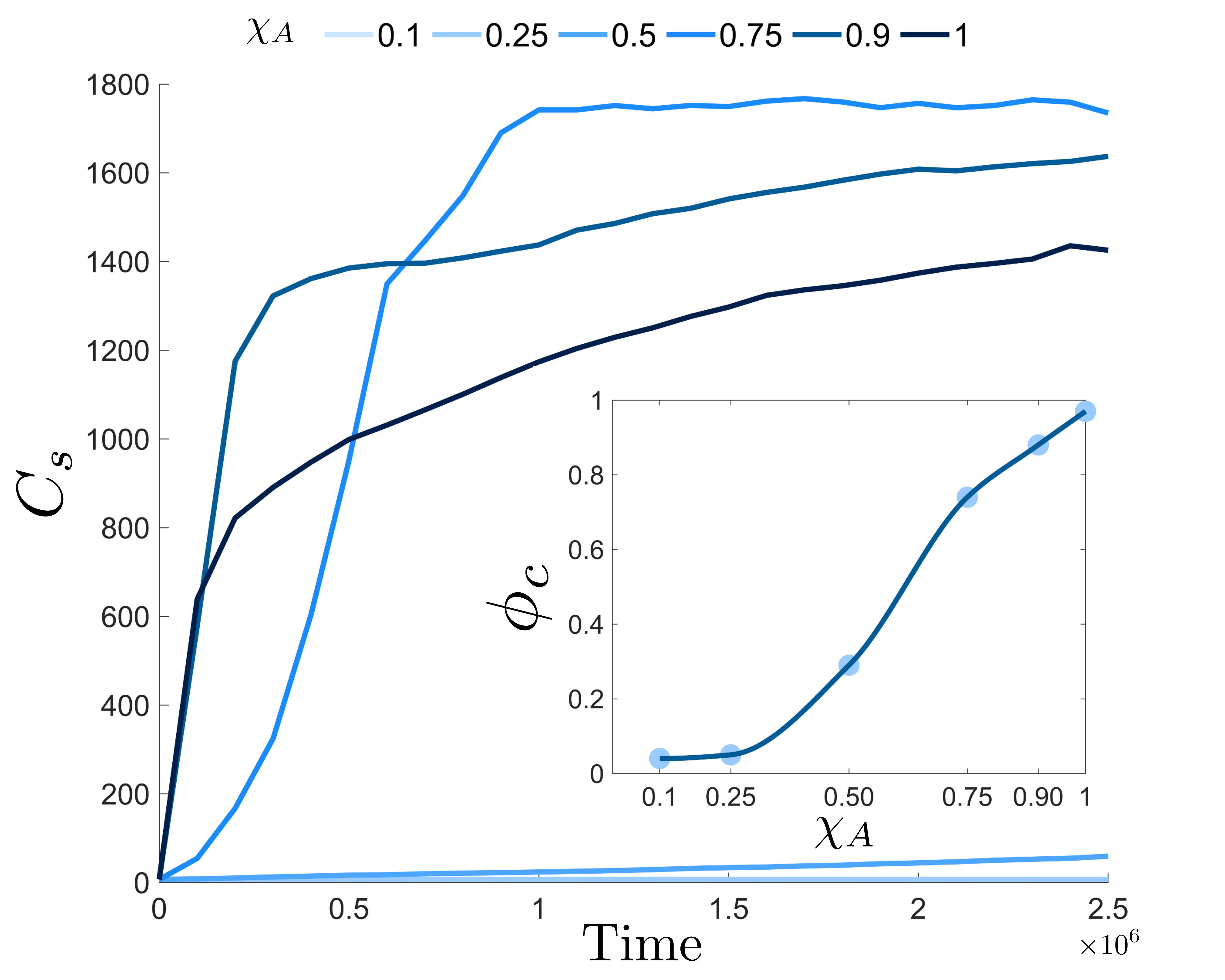}
    \caption{Cluster analysis for different fractions $\chi_A$ of active particles at fixed $\rho=0.05$. Cluster size and fraction of particles in clusters as in Fig. \ref{cluster} for different $\chi_A$ in a binary mixture of active and passive crowders.
    }
    \label{cluster-mix}
\end{figure}

 Finally,  we consider a binary mixture of active and passive crowders,  with $\chi_A +\chi_P=1$ where $\chi_P$ is the fraction of passive particles. We tracked the effect of the three control parameters $\chi_A$, $\alpha$ and $\rho$ on the diffusion coefficient of a tracer particle with $\beta =0.9$. Adding passive particles into a purely active system changes not only the directionality imposed by collisions (due to pushing by the active particles), but also alters the global dynamics in a subtle way, specifically the dynamics of  clustering. In the example shown in Fig.~\ref{cluster-mix} with a fixed low density of crowders, an increase of the fraction of active crowders above 0.5 leads to a strong clustering signature both in the average cluster size $C_s$ and in the fraction $\phi_c$ of particles within clusters (inset). The cluster size $C_s$ shows biphasic growth for large fractions of active crowders with first rapid increase in the cluster size and then a much slower second phase, which appears to be more pronounced and slower for the largest active crowder fractions. 
Put differently,  replacing some of the active crowders with passive particles delays the phase separation (compare the cases $\chi_A=1$ and $\chi_A=0.75$ in Fig.~\ref{cluster-mix}), likely because initially the rapid directional change of the passive particles randomizes the motion of the active particles. Eventually, however, clustering sets in. While clustering sets in later for $\chi_A=0.75$ than for $\chi_A=1$, it reaches its stationary state faster. 

\begin{figure*}[t]
   \centering
   \includegraphics[width=1.\textwidth]{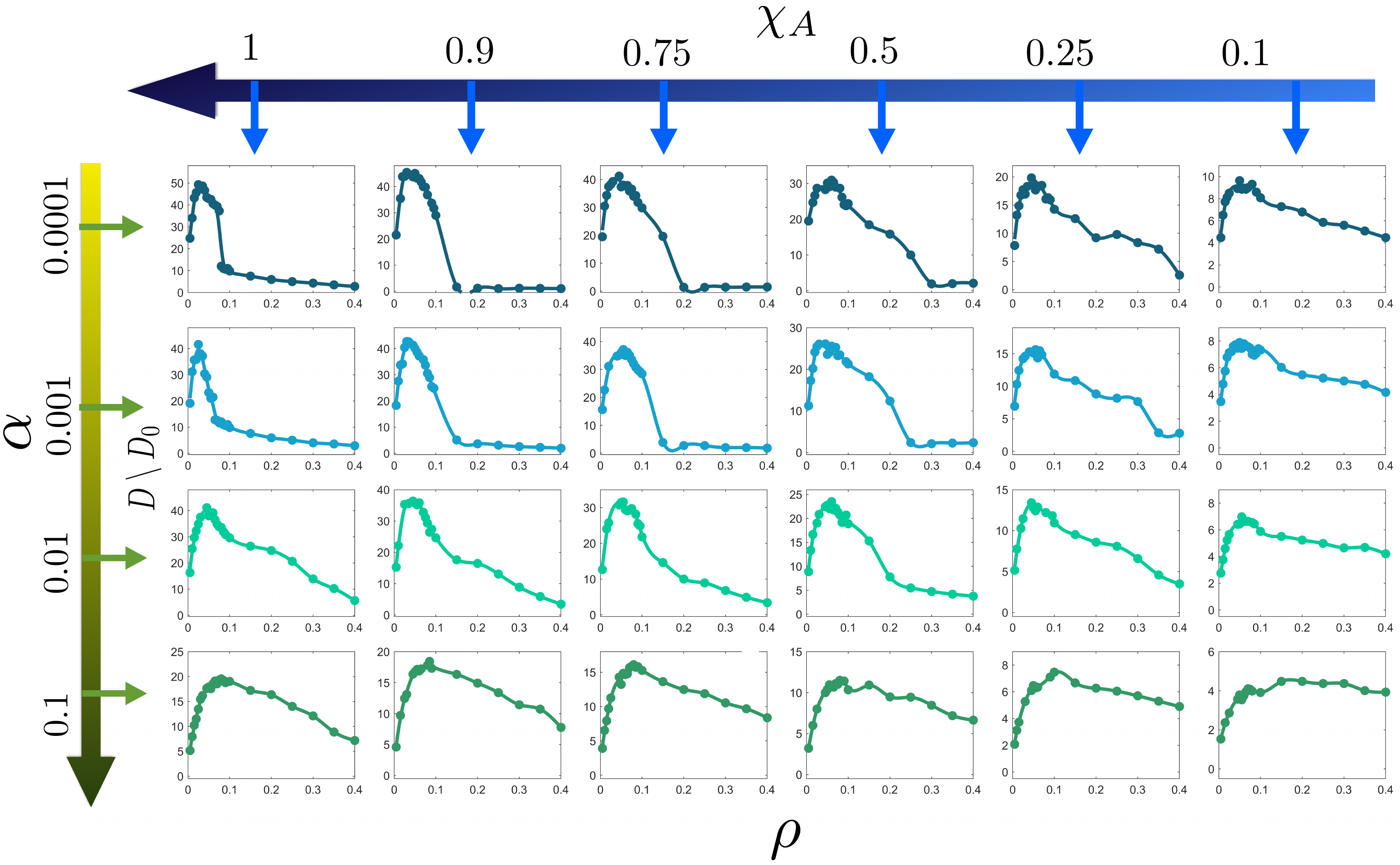}
    \caption{The diffusion coefficient of a tracer with $\beta=0.9$  as a function of the density $\rho$ and of the fraction of active crowders $\chi_A$ in a binary mixture. }
    \label{Mix}
\end{figure*}

Above we have shown a correspondence between the onset of clustering and a decrease of the enhanced diffusion of a passive tracer (Fig.~\ref{Diffusion}). Since clustering is decreased by increasing the fraction of passive particles in the system, we expect that the decrease in diffusion of the tracer is less pronounced when passive particles are present. This is indeed the case, as shown in Fig.~\ref{Mix}, where we plot the diffusion coefficient as a function of the density for different combinations of the tumbling probability $\alpha$ and the fraction of active crowders $\chi_A$: decreasing the fraction of active crowders leads to a broader peak in the diffusion coefficient as a function of density. We interpret this observation as reflecting the fact that a larger fraction of the active particles is in the gas phase and can push the tracer. At the same time, however, the value of the maximal diffusion coefficient is also decreased, reflecting the fact that the total number of active particles is reduced. 
Nevertheless, even for very low fractions of active particles (bottom row in Fig.~\ref{Mix}), the diffusion coefficient is still increased compared to the case with only passive particles. 
For instance, for $\alpha=0.1$, $\chi_A= 0.1$, it  is increased up to $5$-fold.


\section{Conclusion}
In this study, we have analyzed the diffusion of a tracer particle in a crowded active system using a minimal lattice model. We implemented the pushing of a passive tracer by active crowder particles by a simple collision rule. A passive tracer in an otherwise purely active system or in a binary mixture of active and passive particles  exhibits different motility regimes including significantly enhanced effective diffusion. Our results show that the extent to which diffusion is enhanced depends on the activity of the crowder particles (modulated via their hopping rate relative to the tracer's hopping rate and via their tumbling probability, i.e. via the persistence of their motion) and on their density. The interplay of activity and density depends on the dynamics of the particles directly, but also indirectly via the phase separation of the crowders into low-density (gas-phase) regions and dense clusters. The latter has a negative effect on the mobility of the tracer, both when the tracer becomes trapped in a cluster and when the tracer moves in a low-density area, where it is not pushed by active particles.  Our findings point to an intricate interplay between the local dynamics of enhanced tracer diffusion and the global dynamics of the system. Given the simplicity of our model, one can extend it to other collision rules among the particles as well as use it study impact of active crowding on other physical processes.

\section*{Acknowledgement}
This research was conducted within the Max Planck School Matter to Life supported by the German Federal Ministry of Education and Research (BMBF) in collaboration with the Max Planck Society. The authors acknowledge further support by DFG through  SFB 937 project A21 in the initial phase of this work.

\bibliography{reference}

\begin{thebibliography}{44}%
\makeatletter
\providecommand \@ifxundefined [1]{%
 \@ifx{#1\undefined}
}%
\providecommand \@ifnum [1]{%
 \ifnum #1\expandafter \@firstoftwo
 \else \expandafter \@secondoftwo
 \fi
}%
\providecommand \@ifx [1]{%
 \ifx #1\expandafter \@firstoftwo
 \else \expandafter \@secondoftwo
 \fi
}%
\providecommand \natexlab [1]{#1}%
\providecommand \enquote  [1]{``#1''}%
\providecommand \bibnamefont  [1]{#1}%
\providecommand \bibfnamefont [1]{#1}%
\providecommand \citenamefont [1]{#1}%
\providecommand \href@noop [0]{\@secondoftwo}%
\providecommand \href [0]{\begingroup \@sanitize@url \@href}%
\providecommand \@href[1]{\@@startlink{#1}\@@href}%
\providecommand \@@href[1]{\endgroup#1\@@endlink}%
\providecommand \@sanitize@url [0]{\catcode `\\12\catcode `\$12\catcode
  `\&12\catcode `\#12\catcode `\^12\catcode `\_12\catcode `\%12\relax}%
\providecommand \@@startlink[1]{}%
\providecommand \@@endlink[0]{}%
\providecommand \url  [0]{\begingroup\@sanitize@url \@url }%
\providecommand \@url [1]{\endgroup\@href {#1}{\urlprefix }}%
\providecommand \urlprefix  [0]{URL }%
\providecommand \Eprint [0]{\href }%
\providecommand \doibase [0]{http://dx.doi.org/}%
\providecommand \selectlanguage [0]{\@gobble}%
\providecommand \bibinfo  [0]{\@secondoftwo}%
\providecommand \bibfield  [0]{\@secondoftwo}%
\providecommand \translation [1]{[#1]}%
\providecommand \BibitemOpen [0]{}%
\providecommand \bibitemStop [0]{}%
\providecommand \bibitemNoStop [0]{.\EOS\space}%
\providecommand \EOS [0]{\spacefactor3000\relax}%
\providecommand \BibitemShut  [1]{\csname bibitem#1\endcsname}%
\let\auto@bib@innerbib\@empty
\bibitem [{\citenamefont {Ellis}(2001)}]{ellis2001}%
  \BibitemOpen
  \bibfield  {author} {\bibinfo {author} {\bibfnamefont {R~John}\ \bibnamefont
  {Ellis}},\ }\bibfield  {title} {\enquote {\bibinfo {title} {Macromolecular
  crowding: obvious but underappreciated},}\ }\href
  {https://doi.org/10.1016/S0968-0004(01)01938-7} {\bibfield  {journal}
  {\bibinfo  {journal} {Trends in biochemical sciences}\ }\textbf {\bibinfo
  {volume} {26}},\ \bibinfo {pages} {597--604} (\bibinfo {year}
  {2001})}\BibitemShut {NoStop}%
\bibitem [{\citenamefont {Dix}\ and\ \citenamefont {Verkman}(2008)}]{Dix2008}%
  \BibitemOpen
  \bibfield  {author} {\bibinfo {author} {\bibfnamefont {James~A.}\
  \bibnamefont {Dix}}\ and\ \bibinfo {author} {\bibfnamefont {A.S.}\
  \bibnamefont {Verkman}},\ }\bibfield  {title} {\enquote {\bibinfo {title}
  {Crowding effects on diffusion in solutions and cells},}\ }\href {\doibase
  10.1146/annurev.biophys.37.032807.125824} {\bibfield  {journal} {\bibinfo
  {journal} {Annual Review of Biophysics}\ }\textbf {\bibinfo {volume} {37}},\
  \bibinfo {pages} {247--263} (\bibinfo {year} {2008})}\BibitemShut {NoStop}%
\bibitem [{\citenamefont {Bechinger}\ \emph {et~al.}(2016)\citenamefont
  {Bechinger}, \citenamefont {Di~Leonardo}, \citenamefont {L\"owen},
  \citenamefont {Reichhardt}, \citenamefont {Volpe},\ and\ \citenamefont
  {Volpe}}]{Bechinger2016}%
  \BibitemOpen
  \bibfield  {author} {\bibinfo {author} {\bibfnamefont {Clemens}\ \bibnamefont
  {Bechinger}}, \bibinfo {author} {\bibfnamefont {Roberto}\ \bibnamefont
  {Di~Leonardo}}, \bibinfo {author} {\bibfnamefont {Hartmut}\ \bibnamefont
  {L\"owen}}, \bibinfo {author} {\bibfnamefont {Charles}\ \bibnamefont
  {Reichhardt}}, \bibinfo {author} {\bibfnamefont {Giorgio}\ \bibnamefont
  {Volpe}}, \ and\ \bibinfo {author} {\bibfnamefont {Giovanni}\ \bibnamefont
  {Volpe}},\ }\bibfield  {title} {\enquote {\bibinfo {title} {Active particles
  in complex and crowded environments},}\ }\href {\doibase
  10.1103/RevModPhys.88.045006} {\bibfield  {journal} {\bibinfo  {journal}
  {Rev. Mod. Phys.}\ }\textbf {\bibinfo {volume} {88}},\ \bibinfo {pages}
  {045006} (\bibinfo {year} {2016})}\BibitemShut {NoStop}%
\bibitem [{\citenamefont {Klumpp}\ \emph {et~al.}(2019)\citenamefont {Klumpp},
  \citenamefont {Bode},\ and\ \citenamefont {Puri}}]{klumpp2019}%
  \BibitemOpen
  \bibfield  {author} {\bibinfo {author} {\bibfnamefont {Stefan}\ \bibnamefont
  {Klumpp}}, \bibinfo {author} {\bibfnamefont {William}\ \bibnamefont {Bode}},
  \ and\ \bibinfo {author} {\bibfnamefont {Palka}\ \bibnamefont {Puri}},\
  }\bibfield  {title} {\enquote {\bibinfo {title} {Life in crowded
  conditions},}\ }\href@noop {} {\bibfield  {journal} {\bibinfo  {journal} {The
  European Physical Journal Special Topics}\ ,\ \bibinfo {pages} {1--14}}
  (\bibinfo {year} {2019})},\ \Eprint
  {http://arxiv.org/abs/https://doi.org/10.1140/ep jst/e2018-800088-6}
  {https://doi.org/10.1140/ep jst/e2018-800088-6} \BibitemShut {NoStop}%
\bibitem [{\citenamefont {Zimmerman}\ and\ \citenamefont
  {Minton}(1993)}]{Zimmerman1993}%
  \BibitemOpen
  \bibfield  {author} {\bibinfo {author} {\bibfnamefont {S~B}\ \bibnamefont
  {Zimmerman}}\ and\ \bibinfo {author} {\bibfnamefont {A~P}\ \bibnamefont
  {Minton}},\ }\bibfield  {title} {\enquote {\bibinfo {title} {Macromolecular
  crowding: Biochemical, biophysical, and physiological consequences},}\ }\href
  {\doibase 10.1146/annurev.bb.22.060193.000331} {\bibfield  {journal}
  {\bibinfo  {journal} {Annual Review of Biophysics and Biomolecular
  Structure}\ }\textbf {\bibinfo {volume} {22}},\ \bibinfo {pages} {27--65}
  (\bibinfo {year} {1993})}\BibitemShut {NoStop}%
\bibitem [{\citenamefont {Mendelson}\ \emph {et~al.}(1999)\citenamefont
  {Mendelson}, \citenamefont {Bourque}, \citenamefont {Wilkening},
  \citenamefont {Anderson},\ and\ \citenamefont {Watkins}}]{Mendelson1999}%
  \BibitemOpen
  \bibfield  {author} {\bibinfo {author} {\bibfnamefont {Neil~H.}\ \bibnamefont
  {Mendelson}}, \bibinfo {author} {\bibfnamefont {Adrienne}\ \bibnamefont
  {Bourque}}, \bibinfo {author} {\bibfnamefont {Kathryn}\ \bibnamefont
  {Wilkening}}, \bibinfo {author} {\bibfnamefont {Kevin~R.}\ \bibnamefont
  {Anderson}}, \ and\ \bibinfo {author} {\bibfnamefont {Joseph~C.}\
  \bibnamefont {Watkins}},\ }\bibfield  {title} {\enquote {\bibinfo {title}
  {Organized cell swimming motions in bacillus subtilis colonies: Patterns of
  short-lived whirls and jets},}\ }\href {https://jb.asm.org/content/181/2/600}
  {\bibfield  {journal} {\bibinfo  {journal} {American Society for Microbiology
  Journals}\ }\textbf {\bibinfo {volume} {181}},\ \bibinfo {pages} {600--609}
  (\bibinfo {year} {1999})}\BibitemShut {NoStop}%
\bibitem [{\citenamefont {Hall-Stoodley}\ \emph {et~al.}(2004)\citenamefont
  {Hall-Stoodley}, \citenamefont {Costerton},\ and\ \citenamefont
  {Stoodley}}]{Hall2004}%
  \BibitemOpen
  \bibfield  {author} {\bibinfo {author} {\bibfnamefont {Luanne}\ \bibnamefont
  {Hall-Stoodley}}, \bibinfo {author} {\bibfnamefont {J~William}\ \bibnamefont
  {Costerton}}, \ and\ \bibinfo {author} {\bibfnamefont {Paul}\ \bibnamefont
  {Stoodley}},\ }\bibfield  {title} {\enquote {\bibinfo {title} {Bacterial
  biofilms: from the natural environment to infectious diseases},}\ }\href@noop
  {} {\bibfield  {journal} {\bibinfo  {journal} {Nature reviews microbiology}\
  }\textbf {\bibinfo {volume} {2}},\ \bibinfo {pages} {95--108} (\bibinfo
  {year} {2004})},\ \Eprint
  {http://arxiv.org/abs/https://doi.org/10.1038/nrmicro821}
  {https://doi.org/10.1038/nrmicro821} \BibitemShut {NoStop}%
\bibitem [{\citenamefont {Salbreux}\ and\ \citenamefont
  {J\"ulicher}(2017)}]{Salbreux2017}%
  \BibitemOpen
  \bibfield  {author} {\bibinfo {author} {\bibfnamefont {Guillaume}\
  \bibnamefont {Salbreux}}\ and\ \bibinfo {author} {\bibfnamefont {Frank}\
  \bibnamefont {J\"ulicher}},\ }\bibfield  {title} {\enquote {\bibinfo {title}
  {Mechanics of active surfaces},}\ }\href {\doibase
  10.1103/PhysRevE.96.032404} {\bibfield  {journal} {\bibinfo  {journal} {Phys.
  Rev. E}\ }\textbf {\bibinfo {volume} {96}},\ \bibinfo {pages} {032404}
  (\bibinfo {year} {2017})}\BibitemShut {NoStop}%
\bibitem [{\citenamefont {Mizuno}\ \emph {et~al.}(2007)\citenamefont {Mizuno},
  \citenamefont {Tardin}, \citenamefont {Schmidt},\ and\ \citenamefont
  {MacKintosh}}]{MacKintosh2007}%
  \BibitemOpen
  \bibfield  {author} {\bibinfo {author} {\bibfnamefont {Daisuke}\ \bibnamefont
  {Mizuno}}, \bibinfo {author} {\bibfnamefont {Catherine}\ \bibnamefont
  {Tardin}}, \bibinfo {author} {\bibfnamefont {Christoph~F}\ \bibnamefont
  {Schmidt}}, \ and\ \bibinfo {author} {\bibfnamefont {Frederik~C}\
  \bibnamefont {MacKintosh}},\ }\bibfield  {title} {\enquote {\bibinfo {title}
  {Nonequilibrium mechanics of active cytoskeletal networks},}\ }\href
  {https://science.sciencemag.org/content/sci/315/5810/370.full.pdf} {\bibfield
   {journal} {\bibinfo  {journal} {Science}\ }\textbf {\bibinfo {volume}
  {315}},\ \bibinfo {pages} {370--373} (\bibinfo {year} {2007})}\BibitemShut
  {NoStop}%
\bibitem [{\citenamefont {Muller}\ \emph {et~al.}(2008)\citenamefont {Muller},
  \citenamefont {Klumpp},\ and\ \citenamefont {Lipowsky}}]{Klumpp2008}%
  \BibitemOpen
  \bibfield  {author} {\bibinfo {author} {\bibfnamefont {M.~J.~I.}\
  \bibnamefont {Muller}}, \bibinfo {author} {\bibfnamefont {S.}~\bibnamefont
  {Klumpp}}, \ and\ \bibinfo {author} {\bibfnamefont {R.}~\bibnamefont
  {Lipowsky}},\ }\bibfield  {title} {\enquote {\bibinfo {title} {Tug-of-war as
  a cooperative mechanism for bidirectional cargo transport by molecular
  motors},}\ }\href {\doibase 10.1073/pnas.0706825105} {\bibfield  {journal}
  {\bibinfo  {journal} {Proceedings of the National Academy of Sciences}\
  }\textbf {\bibinfo {volume} {105}},\ \bibinfo {pages} {4609--4614} (\bibinfo
  {year} {2008})}\BibitemShut {NoStop}%
\bibitem [{\citenamefont {Korn}\ \emph {et~al.}(2009)\citenamefont {Korn},
  \citenamefont {Klumpp}, \citenamefont {Lipowsky},\ and\ \citenamefont
  {Schwarz}}]{Korn2009}%
  \BibitemOpen
  \bibfield  {author} {\bibinfo {author} {\bibfnamefont {Christian~B.}\
  \bibnamefont {Korn}}, \bibinfo {author} {\bibfnamefont {Stefan}\ \bibnamefont
  {Klumpp}}, \bibinfo {author} {\bibfnamefont {Reinhard}\ \bibnamefont
  {Lipowsky}}, \ and\ \bibinfo {author} {\bibfnamefont {Ulrich~S.}\
  \bibnamefont {Schwarz}},\ }\bibfield  {title} {\enquote {\bibinfo {title}
  {Stochastic simulations of cargo transport by processive molecular motors},}\
  }\href {\doibase 10.1063/1.3279305} {\bibfield  {journal} {\bibinfo
  {journal} {The Journal of Chemical Physics}\ }\textbf {\bibinfo {volume}
  {131}},\ \bibinfo {pages} {245107} (\bibinfo {year} {2009})},\ \Eprint
  {http://arxiv.org/abs/https://doi.org/10.1063/1.3279305}
  {https://doi.org/10.1063/1.3279305} \BibitemShut {NoStop}%
\bibitem [{\citenamefont {Golestanian}(2009)}]{Ramin2009}%
  \BibitemOpen
  \bibfield  {author} {\bibinfo {author} {\bibfnamefont {Ramin}\ \bibnamefont
  {Golestanian}},\ }\bibfield  {title} {\enquote {\bibinfo {title} {Anomalous
  diffusion of symmetric and asymmetric active colloids},}\ }\href {\doibase
  10.1103/physrevlett.102.188305} {\bibfield  {journal} {\bibinfo  {journal}
  {Physical Review Letters}\ }\textbf {\bibinfo {volume} {102}} (\bibinfo
  {year} {2009}),\ 10.1103/physrevlett.102.188305}\BibitemShut {NoStop}%
\bibitem [{\citenamefont {Illien}\ \emph {et~al.}(2017)\citenamefont {Illien},
  \citenamefont {Adeleke-Larodo},\ and\ \citenamefont
  {Golestanian}}]{Illien2017}%
  \BibitemOpen
  \bibfield  {author} {\bibinfo {author} {\bibfnamefont {P.}~\bibnamefont
  {Illien}}, \bibinfo {author} {\bibfnamefont {T.}~\bibnamefont
  {Adeleke-Larodo}}, \ and\ \bibinfo {author} {\bibfnamefont {R.}~\bibnamefont
  {Golestanian}},\ }\bibfield  {title} {\enquote {\bibinfo {title} {Diffusion
  of an enzyme: The role of fluctuation-induced hydrodynamic coupling},}\
  }\href {\doibase 10.1209/0295-5075/119/40002} {\bibfield  {journal} {\bibinfo
   {journal} {{EPL} (Europhysics Letters)}\ }\textbf {\bibinfo {volume}
  {119}},\ \bibinfo {pages} {40002} (\bibinfo {year} {2017})}\BibitemShut
  {NoStop}%
\bibitem [{\citenamefont {Sokolov}\ and\ \citenamefont
  {Aranson}(2012)}]{Sokolov2012}%
  \BibitemOpen
  \bibfield  {author} {\bibinfo {author} {\bibfnamefont {Andrey}\ \bibnamefont
  {Sokolov}}\ and\ \bibinfo {author} {\bibfnamefont {Igor~S.}\ \bibnamefont
  {Aranson}},\ }\bibfield  {title} {\enquote {\bibinfo {title} {Physical
  properties of collective motion in suspensions of bacteria},}\ }\href
  {\doibase 10.1103/PhysRevLett.109.248109} {\bibfield  {journal} {\bibinfo
  {journal} {Phys. Rev. Lett.}\ }\textbf {\bibinfo {volume} {109}},\ \bibinfo
  {pages} {248109} (\bibinfo {year} {2012})}\BibitemShut {NoStop}%
\bibitem [{\citenamefont {Dunkel}\ \emph {et~al.}(2013)\citenamefont {Dunkel},
  \citenamefont {Heidenreich}, \citenamefont {Drescher}, \citenamefont
  {Wensink}, \citenamefont {B\"ar},\ and\ \citenamefont
  {Goldstein}}]{Dunkel2013}%
  \BibitemOpen
  \bibfield  {author} {\bibinfo {author} {\bibfnamefont {J\"orn}\ \bibnamefont
  {Dunkel}}, \bibinfo {author} {\bibfnamefont {Sebastian}\ \bibnamefont
  {Heidenreich}}, \bibinfo {author} {\bibfnamefont {Knut}\ \bibnamefont
  {Drescher}}, \bibinfo {author} {\bibfnamefont {Henricus~H.}\ \bibnamefont
  {Wensink}}, \bibinfo {author} {\bibfnamefont {Markus}\ \bibnamefont {B\"ar}},
  \ and\ \bibinfo {author} {\bibfnamefont {Raymond~E.}\ \bibnamefont
  {Goldstein}},\ }\bibfield  {title} {\enquote {\bibinfo {title} {Fluid
  dynamics of bacterial turbulence},}\ }\href {\doibase
  10.1103/PhysRevLett.110.228102} {\bibfield  {journal} {\bibinfo  {journal}
  {Phys. Rev. Lett.}\ }\textbf {\bibinfo {volume} {110}},\ \bibinfo {pages}
  {228102} (\bibinfo {year} {2013})}\BibitemShut {NoStop}%
\bibitem [{\citenamefont {Konopka}\ \emph {et~al.}(2009)\citenamefont
  {Konopka}, \citenamefont {Sochacki}, \citenamefont {Bratton}, \citenamefont
  {Shkel}, \citenamefont {Record},\ and\ \citenamefont
  {Weisshaar}}]{konopka2009}%
  \BibitemOpen
  \bibfield  {author} {\bibinfo {author} {\bibfnamefont {Michael~C}\
  \bibnamefont {Konopka}}, \bibinfo {author} {\bibfnamefont {Kem~A}\
  \bibnamefont {Sochacki}}, \bibinfo {author} {\bibfnamefont {Benjamin~P}\
  \bibnamefont {Bratton}}, \bibinfo {author} {\bibfnamefont {Irina~A}\
  \bibnamefont {Shkel}}, \bibinfo {author} {\bibfnamefont {M~Thomas}\
  \bibnamefont {Record}}, \ and\ \bibinfo {author} {\bibfnamefont {James~C}\
  \bibnamefont {Weisshaar}},\ }\bibfield  {title} {\enquote {\bibinfo {title}
  {Cytoplasmic protein mobility in osmotically stressed escherichia coli},}\
  }\href {https://jb.asm.org/content/191/1/231} {\bibfield  {journal} {\bibinfo
   {journal} {Journal of bacteriology}\ }\textbf {\bibinfo {volume} {191}},\
  \bibinfo {pages} {231--237} (\bibinfo {year} {2009})}\BibitemShut {NoStop}%
\bibitem [{\citenamefont {Parry}\ \emph {et~al.}(2014)\citenamefont {Parry},
  \citenamefont {Surovtsev}, \citenamefont {Cabeen}, \citenamefont {O’Hern},
  \citenamefont {Dufresne},\ and\ \citenamefont {Jacobs-Wagner}}]{parry2014}%
  \BibitemOpen
  \bibfield  {author} {\bibinfo {author} {\bibfnamefont {Bradley~R}\
  \bibnamefont {Parry}}, \bibinfo {author} {\bibfnamefont {Ivan~V}\
  \bibnamefont {Surovtsev}}, \bibinfo {author} {\bibfnamefont {Matthew~T}\
  \bibnamefont {Cabeen}}, \bibinfo {author} {\bibfnamefont {Corey~S}\
  \bibnamefont {O’Hern}}, \bibinfo {author} {\bibfnamefont {Eric~R}\
  \bibnamefont {Dufresne}}, \ and\ \bibinfo {author} {\bibfnamefont
  {Christine}\ \bibnamefont {Jacobs-Wagner}},\ }\bibfield  {title} {\enquote
  {\bibinfo {title} {The bacterial cytoplasm has glass-like properties and is
  fluidized by metabolic activity},}\ }\href
  {https://doi.org/10.1016/j.cell.2013.11.028} {\bibfield  {journal} {\bibinfo
  {journal} {Cell}\ }\textbf {\bibinfo {volume} {156}},\ \bibinfo {pages}
  {183--194} (\bibinfo {year} {2014})}\BibitemShut {NoStop}%
\bibitem [{\citenamefont {Boersma}\ \emph {et~al.}(2015)\citenamefont
  {Boersma}, \citenamefont {Zuhorn},\ and\ \citenamefont
  {Poolman}}]{boersma2015}%
  \BibitemOpen
  \bibfield  {author} {\bibinfo {author} {\bibfnamefont {Arnold~J}\
  \bibnamefont {Boersma}}, \bibinfo {author} {\bibfnamefont {Inge~S}\
  \bibnamefont {Zuhorn}}, \ and\ \bibinfo {author} {\bibfnamefont {Bert}\
  \bibnamefont {Poolman}},\ }\bibfield  {title} {\enquote {\bibinfo {title} {A
  sensor for quantification of macromolecular crowding in living cells},}\
  }\href {https://doi.org/10.1038/nmeth.3257} {\bibfield  {journal} {\bibinfo
  {journal} {Nature methods}\ }\textbf {\bibinfo {volume} {12}},\ \bibinfo
  {pages} {227} (\bibinfo {year} {2015})}\BibitemShut {NoStop}%
\bibitem [{\citenamefont {Gnutt}\ \emph {et~al.}(2015)\citenamefont {Gnutt},
  \citenamefont {Gao}, \citenamefont {Brylski}, \citenamefont {Heyden},\ and\
  \citenamefont {Ebbinghaus}}]{gnutt2015}%
  \BibitemOpen
  \bibfield  {author} {\bibinfo {author} {\bibfnamefont {David}\ \bibnamefont
  {Gnutt}}, \bibinfo {author} {\bibfnamefont {Mimi}\ \bibnamefont {Gao}},
  \bibinfo {author} {\bibfnamefont {Oliver}\ \bibnamefont {Brylski}}, \bibinfo
  {author} {\bibfnamefont {Matthias}\ \bibnamefont {Heyden}}, \ and\ \bibinfo
  {author} {\bibfnamefont {Simon}\ \bibnamefont {Ebbinghaus}},\ }\bibfield
  {title} {\enquote {\bibinfo {title} {Excluded-volume effects in living
  cells},}\ }\href {https://doi.org/10.1002/anie.201409847} {\bibfield
  {journal} {\bibinfo  {journal} {Angewandte Chemie International Edition}\
  }\textbf {\bibinfo {volume} {54}},\ \bibinfo {pages} {2548--2551} (\bibinfo
  {year} {2015})}\BibitemShut {NoStop}%
\bibitem [{\citenamefont {Koch}\ and\ \citenamefont {Brady}(1988)}]{koch1988}%
  \BibitemOpen
  \bibfield  {author} {\bibinfo {author} {\bibfnamefont {Donald~L}\
  \bibnamefont {Koch}}\ and\ \bibinfo {author} {\bibfnamefont {John~F}\
  \bibnamefont {Brady}},\ }\bibfield  {title} {\enquote {\bibinfo {title}
  {Anomalous diffusion in heterogeneous porous media},}\ }\href
  {https://doi.org/10.1063/1.866716} {\bibfield  {journal} {\bibinfo  {journal}
  {The Physics of fluids}\ }\textbf {\bibinfo {volume} {31}},\ \bibinfo {pages}
  {965--973} (\bibinfo {year} {1988})}\BibitemShut {NoStop}%
\bibitem [{\citenamefont {Ghosh}\ \emph {et~al.}(2016)\citenamefont {Ghosh},
  \citenamefont {Cherstvy}, \citenamefont {Grebenkov},\ and\ \citenamefont
  {Metzler}}]{Ghosh2016}%
  \BibitemOpen
  \bibfield  {author} {\bibinfo {author} {\bibfnamefont {Surya~K}\ \bibnamefont
  {Ghosh}}, \bibinfo {author} {\bibfnamefont {Andrey~G}\ \bibnamefont
  {Cherstvy}}, \bibinfo {author} {\bibfnamefont {Denis~S}\ \bibnamefont
  {Grebenkov}}, \ and\ \bibinfo {author} {\bibfnamefont {Ralf}\ \bibnamefont
  {Metzler}},\ }\bibfield  {title} {\enquote {\bibinfo {title} {Anomalous,
  non-gaussian tracer diffusion in crowded two-dimensional environments},}\
  }\href {\doibase 10.1088/1367-2630/18/1/013027} {\bibfield  {journal}
  {\bibinfo  {journal} {New Journal of Physics}\ }\textbf {\bibinfo {volume}
  {18}},\ \bibinfo {pages} {013027} (\bibinfo {year} {2016})}\BibitemShut
  {NoStop}%
\bibitem [{\citenamefont {Wu}\ and\ \citenamefont {Libchaber}(2000)}]{Wu2000}%
  \BibitemOpen
  \bibfield  {author} {\bibinfo {author} {\bibfnamefont {Xiao-Lun}\
  \bibnamefont {Wu}}\ and\ \bibinfo {author} {\bibfnamefont {Albert}\
  \bibnamefont {Libchaber}},\ }\bibfield  {title} {\enquote {\bibinfo {title}
  {Particle diffusion in a quasi-two-dimensional bacterial bath},}\ }\href
  {\doibase 10.1103/PhysRevLett.84.3017} {\bibfield  {journal} {\bibinfo
  {journal} {Phys. Rev. Lett.}\ }\textbf {\bibinfo {volume} {84}},\ \bibinfo
  {pages} {3017--3020} (\bibinfo {year} {2000})}\BibitemShut {NoStop}%
\bibitem [{\citenamefont {Gr\'egoire}\ \emph {et~al.}(2001)\citenamefont
  {Gr\'egoire}, \citenamefont {Chat\'e},\ and\ \citenamefont
  {Tu}}]{Hugues2001}%
  \BibitemOpen
  \bibfield  {author} {\bibinfo {author} {\bibfnamefont {Guillaume}\
  \bibnamefont {Gr\'egoire}}, \bibinfo {author} {\bibfnamefont {Hugues}\
  \bibnamefont {Chat\'e}}, \ and\ \bibinfo {author} {\bibfnamefont {Yuhai}\
  \bibnamefont {Tu}},\ }\bibfield  {title} {\enquote {\bibinfo {title} {Active
  and passive particles: Modeling beads in a bacterial bath},}\ }\href
  {\doibase 10.1103/PhysRevE.64.011902} {\bibfield  {journal} {\bibinfo
  {journal} {Phys. Rev. E}\ }\textbf {\bibinfo {volume} {64}},\ \bibinfo
  {pages} {011902} (\bibinfo {year} {2001})}\BibitemShut {NoStop}%
\bibitem [{\citenamefont {Zaid}\ \emph {et~al.}(2011)\citenamefont {Zaid},
  \citenamefont {Dunkel},\ and\ \citenamefont {Yeomans}}]{Zaid2011}%
  \BibitemOpen
  \bibfield  {author} {\bibinfo {author} {\bibfnamefont {Irwin~M}\ \bibnamefont
  {Zaid}}, \bibinfo {author} {\bibfnamefont {Jörn}\ \bibnamefont {Dunkel}}, \
  and\ \bibinfo {author} {\bibfnamefont {Julia~M}\ \bibnamefont {Yeomans}},\
  }\bibfield  {title} {\enquote {\bibinfo {title} {Lévy fluctuations and
  mixing in dilute suspensions of algae and bacteria},}\ }\href {\doibase
  10.1098/rsif.2010.0545} {\bibfield  {journal} {\bibinfo  {journal} {Journal
  of the Royal Society, Interface}\ }\textbf {\bibinfo {volume} {8}},\ \bibinfo
  {pages} {1314—1331} (\bibinfo {year} {2011})},\ \Eprint
  {http://arxiv.org/abs/https://doi.org/10.1098/rsif.2010.0545}
  {https://doi.org/10.1098/rsif.2010.0545} \BibitemShut {NoStop}%
\bibitem [{\citenamefont {Valeriani}\ \emph {et~al.}(2011)\citenamefont
  {Valeriani}, \citenamefont {Li}, \citenamefont {Novosel}, \citenamefont
  {Arlt},\ and\ \citenamefont {Marenduzzo}}]{valeriani2011}%
  \BibitemOpen
  \bibfield  {author} {\bibinfo {author} {\bibfnamefont {Chantal}\ \bibnamefont
  {Valeriani}}, \bibinfo {author} {\bibfnamefont {Martin}\ \bibnamefont {Li}},
  \bibinfo {author} {\bibfnamefont {John}\ \bibnamefont {Novosel}}, \bibinfo
  {author} {\bibfnamefont {Jochen}\ \bibnamefont {Arlt}}, \ and\ \bibinfo
  {author} {\bibfnamefont {Davide}\ \bibnamefont {Marenduzzo}},\ }\bibfield
  {title} {\enquote {\bibinfo {title} {Colloids in a bacterial bath:
  simulations and experiments},}\ }\href {\doibase 10.1039/C1SM05260H}
  {\bibfield  {journal} {\bibinfo  {journal} {Soft Matter}\ }\textbf {\bibinfo
  {volume} {7}},\ \bibinfo {pages} {5228--5238} (\bibinfo {year}
  {2011})}\BibitemShut {NoStop}%
\bibitem [{\citenamefont {Pushkin}\ and\ \citenamefont
  {Yeomans}(2013)}]{Pushkin2013}%
  \BibitemOpen
  \bibfield  {author} {\bibinfo {author} {\bibfnamefont {Dmitri~O.}\
  \bibnamefont {Pushkin}}\ and\ \bibinfo {author} {\bibfnamefont {Julia~M.}\
  \bibnamefont {Yeomans}},\ }\bibfield  {title} {\enquote {\bibinfo {title}
  {Fluid mixing by curved trajectories of microswimmers},}\ }\href {\doibase
  10.1103/PhysRevLett.111.188101} {\bibfield  {journal} {\bibinfo  {journal}
  {Phys. Rev. Lett.}\ }\textbf {\bibinfo {volume} {111}},\ \bibinfo {pages}
  {188101} (\bibinfo {year} {2013})}\BibitemShut {NoStop}%
\bibitem [{\citenamefont {Soto}\ and\ \citenamefont
  {Golestanian}(2014)}]{Ramin2014}%
  \BibitemOpen
  \bibfield  {author} {\bibinfo {author} {\bibfnamefont {Rodrigo}\ \bibnamefont
  {Soto}}\ and\ \bibinfo {author} {\bibfnamefont {Ramin}\ \bibnamefont
  {Golestanian}},\ }\bibfield  {title} {\enquote {\bibinfo {title}
  {Run-and-tumble dynamics in a crowded environment: Persistent exclusion
  process for swimmers},}\ }\href {\doibase 10.1103/PhysRevE.89.012706}
  {\bibfield  {journal} {\bibinfo  {journal} {Phys. Rev. E}\ }\textbf {\bibinfo
  {volume} {89}},\ \bibinfo {pages} {012706} (\bibinfo {year}
  {2014})}\BibitemShut {NoStop}%
\bibitem [{\citenamefont {Kasyap}\ \emph {et~al.}(2014)\citenamefont {Kasyap},
  \citenamefont {Koch},\ and\ \citenamefont {Wu}}]{Kasyap2014}%
  \BibitemOpen
  \bibfield  {author} {\bibinfo {author} {\bibfnamefont {TV}~\bibnamefont
  {Kasyap}}, \bibinfo {author} {\bibfnamefont {Donald~L}\ \bibnamefont {Koch}},
  \ and\ \bibinfo {author} {\bibfnamefont {Mingming}\ \bibnamefont {Wu}},\
  }\bibfield  {title} {\enquote {\bibinfo {title} {Hydrodynamic tracer
  diffusion in suspensions of swimming bacteria},}\ }\href@noop {} {\bibfield
  {journal} {\bibinfo  {journal} {Physics of Fluids}\ }\textbf {\bibinfo
  {volume} {26}},\ \bibinfo {pages} {081901} (\bibinfo {year} {2014})},\
  \Eprint {http://arxiv.org/abs/https://doi.org/10.1063/1.4891570}
  {https://doi.org/10.1063/1.4891570} \BibitemShut {NoStop}%
\bibitem [{\citenamefont {Morozov}\ and\ \citenamefont
  {Marenduzzo}(2014)}]{Morozov2014}%
  \BibitemOpen
  \bibfield  {author} {\bibinfo {author} {\bibfnamefont {Alexander}\
  \bibnamefont {Morozov}}\ and\ \bibinfo {author} {\bibfnamefont {Davide}\
  \bibnamefont {Marenduzzo}},\ }\bibfield  {title} {\enquote {\bibinfo {title}
  {Enhanced diffusion of tracer particles in dilute bacterial suspensions},}\
  }\href {\doibase 10.1039/C3SM52201F} {\bibfield  {journal} {\bibinfo
  {journal} {Soft Matter}\ }\textbf {\bibinfo {volume} {10}},\ \bibinfo {pages}
  {2748--2758} (\bibinfo {year} {2014})}\BibitemShut {NoStop}%
\bibitem [{\citenamefont {Yang}\ and\ \citenamefont {Bevan}(2020)}]{Yang2020}%
  \BibitemOpen
  \bibfield  {author} {\bibinfo {author} {\bibfnamefont {Yuguang}\ \bibnamefont
  {Yang}}\ and\ \bibinfo {author} {\bibfnamefont {Michael~A.}\ \bibnamefont
  {Bevan}},\ }\bibfield  {title} {\enquote {\bibinfo {title} {Cargo capture and
  transport by colloidal swarms},}\ }\href {\doibase 10.1126/sciadv.aay7679}
  {\bibfield  {journal} {\bibinfo  {journal} {Science Advances}\ }\textbf
  {\bibinfo {volume} {6}} (\bibinfo {year} {2020}),\
  10.1126/sciadv.aay7679}\BibitemShut {NoStop}%
\bibitem [{\citenamefont {Caspi}\ \emph {et~al.}(2000)\citenamefont {Caspi},
  \citenamefont {Granek},\ and\ \citenamefont {Elbaum}}]{Caspi2000}%
  \BibitemOpen
  \bibfield  {author} {\bibinfo {author} {\bibfnamefont {Avi}\ \bibnamefont
  {Caspi}}, \bibinfo {author} {\bibfnamefont {Rony}\ \bibnamefont {Granek}}, \
  and\ \bibinfo {author} {\bibfnamefont {Michael}\ \bibnamefont {Elbaum}},\
  }\bibfield  {title} {\enquote {\bibinfo {title} {Enhanced diffusion in active
  intracellular transport},}\ }\href {\doibase 10.1103/PhysRevLett.85.5655}
  {\bibfield  {journal} {\bibinfo  {journal} {Phys. Rev. Lett.}\ }\textbf
  {\bibinfo {volume} {85}},\ \bibinfo {pages} {5655--5658} (\bibinfo {year}
  {2000})}\BibitemShut {NoStop}%
\bibitem [{\citenamefont {Kim}\ and\ \citenamefont {Breuer}(2004)}]{Kim2004}%
  \BibitemOpen
  \bibfield  {author} {\bibinfo {author} {\bibfnamefont {Min~Jun}\ \bibnamefont
  {Kim}}\ and\ \bibinfo {author} {\bibfnamefont {Kenneth~S.}\ \bibnamefont
  {Breuer}},\ }\bibfield  {title} {\enquote {\bibinfo {title} {Enhanced
  diffusion due to motile bacteria},}\ }\href {\doibase 10.1063/1.1787527}
  {\bibfield  {journal} {\bibinfo  {journal} {Physics of Fluids}\ }\textbf
  {\bibinfo {volume} {16}},\ \bibinfo {pages} {L78--L81} (\bibinfo {year}
  {2004})},\ \Eprint {http://arxiv.org/abs/https://doi.org/10.1063/1.1787527}
  {https://doi.org/10.1063/1.1787527} \BibitemShut {NoStop}%
\bibitem [{\citenamefont {Leptos}\ \emph {et~al.}(2009)\citenamefont {Leptos},
  \citenamefont {Guasto}, \citenamefont {Gollub}, \citenamefont {Pesci},\ and\
  \citenamefont {Goldstein}}]{Leptos2009}%
  \BibitemOpen
  \bibfield  {author} {\bibinfo {author} {\bibfnamefont {Kyriacos~C.}\
  \bibnamefont {Leptos}}, \bibinfo {author} {\bibfnamefont {Jeffrey~S.}\
  \bibnamefont {Guasto}}, \bibinfo {author} {\bibfnamefont {J.~P.}\
  \bibnamefont {Gollub}}, \bibinfo {author} {\bibfnamefont {Adriana~I.}\
  \bibnamefont {Pesci}}, \ and\ \bibinfo {author} {\bibfnamefont {Raymond~E.}\
  \bibnamefont {Goldstein}},\ }\bibfield  {title} {\enquote {\bibinfo {title}
  {Dynamics of enhanced tracer diffusion in suspensions of swimming eukaryotic
  microorganisms},}\ }\href {\doibase 10.1103/PhysRevLett.103.198103}
  {\bibfield  {journal} {\bibinfo  {journal} {Phys. Rev. Lett.}\ }\textbf
  {\bibinfo {volume} {103}},\ \bibinfo {pages} {198103} (\bibinfo {year}
  {2009})}\BibitemShut {NoStop}%
\bibitem [{\citenamefont {Mi\~no}\ \emph {et~al.}(2011)\citenamefont {Mi\~no},
  \citenamefont {Mallouk}, \citenamefont {Darnige}, \citenamefont {Hoyos},
  \citenamefont {Dauchet}, \citenamefont {Dunstan}, \citenamefont {Soto},
  \citenamefont {Wang}, \citenamefont {Rousselet},\ and\ \citenamefont
  {Clement}}]{Mino2011}%
  \BibitemOpen
  \bibfield  {author} {\bibinfo {author} {\bibfnamefont {Gast\'on}\
  \bibnamefont {Mi\~no}}, \bibinfo {author} {\bibfnamefont {Thomas~E.}\
  \bibnamefont {Mallouk}}, \bibinfo {author} {\bibfnamefont {Thierry}\
  \bibnamefont {Darnige}}, \bibinfo {author} {\bibfnamefont {Mauricio}\
  \bibnamefont {Hoyos}}, \bibinfo {author} {\bibfnamefont {Jeremi}\
  \bibnamefont {Dauchet}}, \bibinfo {author} {\bibfnamefont {Jocelyn}\
  \bibnamefont {Dunstan}}, \bibinfo {author} {\bibfnamefont {Rodrigo}\
  \bibnamefont {Soto}}, \bibinfo {author} {\bibfnamefont {Yang}\ \bibnamefont
  {Wang}}, \bibinfo {author} {\bibfnamefont {Annie}\ \bibnamefont {Rousselet}},
  \ and\ \bibinfo {author} {\bibfnamefont {Eric}\ \bibnamefont {Clement}},\
  }\bibfield  {title} {\enquote {\bibinfo {title} {Enhanced diffusion due to
  active swimmers at a solid surface},}\ }\href {\doibase
  10.1103/PhysRevLett.106.048102} {\bibfield  {journal} {\bibinfo  {journal}
  {Phys. Rev. Lett.}\ }\textbf {\bibinfo {volume} {106}},\ \bibinfo {pages}
  {048102} (\bibinfo {year} {2011})}\BibitemShut {NoStop}%
\bibitem [{\citenamefont {Miño}\ \emph {et~al.}(2013)\citenamefont {Miño},
  \citenamefont {Dunstan}, \citenamefont {Rousselet}, \citenamefont
  {Clément},\ and\ \citenamefont {Soto}}]{Mino2013}%
  \BibitemOpen
  \bibfield  {author} {\bibinfo {author} {\bibfnamefont {G.~L.}\ \bibnamefont
  {Miño}}, \bibinfo {author} {\bibfnamefont {J.}~\bibnamefont {Dunstan}},
  \bibinfo {author} {\bibfnamefont {A.}~\bibnamefont {Rousselet}}, \bibinfo
  {author} {\bibfnamefont {E.}~\bibnamefont {Clément}}, \ and\ \bibinfo
  {author} {\bibfnamefont {R.}~\bibnamefont {Soto}},\ }\bibfield  {title}
  {\enquote {\bibinfo {title} {Induced diffusion of tracers in a bacterial
  suspension: theory and experiments},}\ }\href {\doibase 10.1017/jfm.2013.304}
  {\bibfield  {journal} {\bibinfo  {journal} {Journal of Fluid Mechanics}\
  }\textbf {\bibinfo {volume} {729}},\ \bibinfo {pages} {423–444} (\bibinfo
  {year} {2013})}\BibitemShut {NoStop}%
\bibitem [{\citenamefont {Jepson}\ \emph {et~al.}(2013)\citenamefont {Jepson},
  \citenamefont {Martinez}, \citenamefont {Schwarz-Linek}, \citenamefont
  {Morozov},\ and\ \citenamefont {Poon}}]{Jepson2013}%
  \BibitemOpen
  \bibfield  {author} {\bibinfo {author} {\bibfnamefont {Alys}\ \bibnamefont
  {Jepson}}, \bibinfo {author} {\bibfnamefont {Vincent~A.}\ \bibnamefont
  {Martinez}}, \bibinfo {author} {\bibfnamefont {Jana}\ \bibnamefont
  {Schwarz-Linek}}, \bibinfo {author} {\bibfnamefont {Alexander}\ \bibnamefont
  {Morozov}}, \ and\ \bibinfo {author} {\bibfnamefont {Wilson C.~K.}\
  \bibnamefont {Poon}},\ }\bibfield  {title} {\enquote {\bibinfo {title}
  {Enhanced diffusion of nonswimmers in a three-dimensional bath of motile
  bacteria},}\ }\href {\doibase 10.1103/PhysRevE.88.041002} {\bibfield
  {journal} {\bibinfo  {journal} {Phys. Rev. E}\ }\textbf {\bibinfo {volume}
  {88}},\ \bibinfo {pages} {041002} (\bibinfo {year} {2013})}\BibitemShut
  {NoStop}%
\bibitem [{\citenamefont {Koumakis}\ \emph {et~al.}(2013)\citenamefont
  {Koumakis}, \citenamefont {Lepore}, \citenamefont {Maggi},\ and\
  \citenamefont {Di~Leonardo}}]{Koumakis2013}%
  \BibitemOpen
  \bibfield  {author} {\bibinfo {author} {\bibfnamefont {N}~\bibnamefont
  {Koumakis}}, \bibinfo {author} {\bibfnamefont {A}~\bibnamefont {Lepore}},
  \bibinfo {author} {\bibfnamefont {C}~\bibnamefont {Maggi}}, \ and\ \bibinfo
  {author} {\bibfnamefont {R}~\bibnamefont {Di~Leonardo}},\ }\bibfield  {title}
  {\enquote {\bibinfo {title} {Targeted delivery of colloids by swimming
  bacteria},}\ }\href@noop {} {\bibfield  {journal} {\bibinfo  {journal}
  {Nature communications}\ }\textbf {\bibinfo {volume} {4}},\ \bibinfo {pages}
  {1--6} (\bibinfo {year} {2013})},\ \Eprint
  {http://arxiv.org/abs/https://doi.org/10.1038/ncomms3588}
  {https://doi.org/10.1038/ncomms3588} \BibitemShut {NoStop}%
\bibitem [{\citenamefont {Angelani}\ \emph {et~al.}(2011)\citenamefont
  {Angelani}, \citenamefont {Maggi}, \citenamefont {Bernardini}, \citenamefont
  {Rizzo},\ and\ \citenamefont {Di~Leonardo}}]{Angelani2011}%
  \BibitemOpen
  \bibfield  {author} {\bibinfo {author} {\bibfnamefont {L.}~\bibnamefont
  {Angelani}}, \bibinfo {author} {\bibfnamefont {C.}~\bibnamefont {Maggi}},
  \bibinfo {author} {\bibfnamefont {M.~L.}\ \bibnamefont {Bernardini}},
  \bibinfo {author} {\bibfnamefont {A.}~\bibnamefont {Rizzo}}, \ and\ \bibinfo
  {author} {\bibfnamefont {R.}~\bibnamefont {Di~Leonardo}},\ }\bibfield
  {title} {\enquote {\bibinfo {title} {Effective interactions between colloidal
  particles suspended in a bath of swimming cells},}\ }\href {\doibase
  10.1103/PhysRevLett.107.138302} {\bibfield  {journal} {\bibinfo  {journal}
  {Phys. Rev. Lett.}\ }\textbf {\bibinfo {volume} {107}},\ \bibinfo {pages}
  {138302} (\bibinfo {year} {2011})}\BibitemShut {NoStop}%
\bibitem [{\citenamefont {Ghosh}\ \emph {et~al.}(2015)\citenamefont {Ghosh},
  \citenamefont {Cherstvy},\ and\ \citenamefont {Metzler}}]{Ghosh2015}%
  \BibitemOpen
  \bibfield  {author} {\bibinfo {author} {\bibfnamefont {Surya~K}\ \bibnamefont
  {Ghosh}}, \bibinfo {author} {\bibfnamefont {Andrey~G}\ \bibnamefont
  {Cherstvy}}, \ and\ \bibinfo {author} {\bibfnamefont {Ralf}\ \bibnamefont
  {Metzler}},\ }\bibfield  {title} {\enquote {\bibinfo {title} {Non-universal
  tracer diffusion in crowded media of non-inert obstacles},}\ }\href
  {https://doi.org/10.1039/C4CP03599B} {\bibfield  {journal} {\bibinfo
  {journal} {Physical Chemistry Chemical Physics}\ }\textbf {\bibinfo {volume}
  {17}},\ \bibinfo {pages} {1847--1858} (\bibinfo {year} {2015})}\BibitemShut
  {NoStop}%
\bibitem [{\citenamefont {B{\'e}nichou}\ \emph {et~al.}(2018)\citenamefont
  {B{\'e}nichou}, \citenamefont {Illien}, \citenamefont {Oshanin},
  \citenamefont {Sarracino},\ and\ \citenamefont {Voituriez}}]{benichou2018}%
  \BibitemOpen
  \bibfield  {author} {\bibinfo {author} {\bibfnamefont {O}~\bibnamefont
  {B{\'e}nichou}}, \bibinfo {author} {\bibfnamefont {P}~\bibnamefont {Illien}},
  \bibinfo {author} {\bibfnamefont {G}~\bibnamefont {Oshanin}}, \bibinfo
  {author} {\bibfnamefont {A}~\bibnamefont {Sarracino}}, \ and\ \bibinfo
  {author} {\bibfnamefont {R}~\bibnamefont {Voituriez}},\ }\bibfield  {title}
  {\enquote {\bibinfo {title} {Tracer diffusion in crowded narrow channels},}\
  }\href {https://doi.org/10.1088%2F1361-648x%2Faae13a} {\bibfield  {journal}
  {\bibinfo  {journal} {Journal of Physics: Condensed Matter}\ }\textbf
  {\bibinfo {volume} {30}},\ \bibinfo {pages} {443001} (\bibinfo {year}
  {2018})}\BibitemShut {NoStop}%
\bibitem [{\citenamefont {Mejia-Monasterio}\ \emph {et~al.}(2020)\citenamefont
  {Mejia-Monasterio}, \citenamefont {Nechaev}, \citenamefont {Oshanin},\ and\
  \citenamefont {Vasilyev}}]{mejia2020}%
  \BibitemOpen
  \bibfield  {author} {\bibinfo {author} {\bibfnamefont {Carlos}\ \bibnamefont
  {Mejia-Monasterio}}, \bibinfo {author} {\bibfnamefont {Sergei}\ \bibnamefont
  {Nechaev}}, \bibinfo {author} {\bibfnamefont {Gleb}\ \bibnamefont {Oshanin}},
  \ and\ \bibinfo {author} {\bibfnamefont {Oleg}\ \bibnamefont {Vasilyev}},\
  }\bibfield  {title} {\enquote {\bibinfo {title} {Tracer diffusion on a
  crowded random manhattan lattice},}\ }\href@noop {} {\bibfield  {journal}
  {\bibinfo  {journal} {New Journal of Physics}\ }\textbf {\bibinfo {volume}
  {22}},\ \bibinfo {pages} {033024} (\bibinfo {year} {2020})}\BibitemShut
  {NoStop}%
\bibitem [{\citenamefont {Burkholder}\ and\ \citenamefont
  {Brady}(2017)}]{brady2017}%
  \BibitemOpen
  \bibfield  {author} {\bibinfo {author} {\bibfnamefont {Eric~W.}\ \bibnamefont
  {Burkholder}}\ and\ \bibinfo {author} {\bibfnamefont {John~F.}\ \bibnamefont
  {Brady}},\ }\bibfield  {title} {\enquote {\bibinfo {title} {Tracer diffusion
  in active suspensions},}\ }\href {\doibase 10.1103/PhysRevE.95.052605}
  {\bibfield  {journal} {\bibinfo  {journal} {Phys. Rev. E}\ }\textbf {\bibinfo
  {volume} {95}},\ \bibinfo {pages} {052605} (\bibinfo {year}
  {2017})}\BibitemShut {NoStop}%
\bibitem [{\citenamefont {Braun}\ and\ \citenamefont
  {Sholl}(1998)}]{braun1998}%
  \BibitemOpen
  \bibfield  {author} {\bibinfo {author} {\bibfnamefont {O.~M.}\ \bibnamefont
  {Braun}}\ and\ \bibinfo {author} {\bibfnamefont {C.~A.}\ \bibnamefont
  {Sholl}},\ }\bibfield  {title} {\enquote {\bibinfo {title} {Diffusion in
  generalized lattice-gas models},}\ }\href {\doibase
  10.1103/PhysRevB.58.14870} {\bibfield  {journal} {\bibinfo  {journal} {Phys.
  Rev. B}\ }\textbf {\bibinfo {volume} {58}},\ \bibinfo {pages} {14870--14879}
  (\bibinfo {year} {1998})}\BibitemShut {NoStop}%
\bibitem [{\citenamefont {Ramasubramani}\ \emph {et~al.}(2019)\citenamefont
  {Ramasubramani}, \citenamefont {Dice}, \citenamefont {Harper}, \citenamefont
  {Spellings}, \citenamefont {Anderson},\ and\ \citenamefont
  {Glotzer}}]{freud}%
  \BibitemOpen
  \bibfield  {author} {\bibinfo {author} {\bibfnamefont {Vyas}\ \bibnamefont
  {Ramasubramani}}, \bibinfo {author} {\bibfnamefont {Bradley~D.}\ \bibnamefont
  {Dice}}, \bibinfo {author} {\bibfnamefont {Eric~S.}\ \bibnamefont {Harper}},
  \bibinfo {author} {\bibfnamefont {Matthew~P.}\ \bibnamefont {Spellings}},
  \bibinfo {author} {\bibfnamefont {Joshua~A.}\ \bibnamefont {Anderson}}, \
  and\ \bibinfo {author} {\bibfnamefont {Sharon~C.}\ \bibnamefont {Glotzer}},\
  }\href@noop {} {\enquote {\bibinfo {title} {freud: A software suite for high
  throughput analysis of particle simulation data},}\ } (\bibinfo {year}
  {2019}),\ \Eprint {http://arxiv.org/abs/arXiv:1906.06317} {arXiv:1906.06317}
  \BibitemShut {NoStop}%
\end{thebibliography}%

\end{document}